\documentclass[aps,prl,nofootinbib,twocolumn,superscriptaddress,preprintnumbers]{revtex4-2}
\usepackage[nodisplayskipstretch]{setspace}
\usepackage{cancel}
\usepackage{mathtools}
\allowdisplaybreaks
\usepackage{tabularx}

\usepackage{amssymb}
\usepackage{amsmath}
\usepackage{epsfig}
\usepackage{hyperref}
\usepackage{breakurl}
\usepackage{bm}
\usepackage{color}
\usepackage{tikz}
\usetikzlibrary{decorations.markings}
\usepackage[english]{babel}
\newcommand\beq{\begin{equation}}
\newcommand\eeq{\end{equation}}
\def\bea{\begin{eqnarray}}
\def\eea{\end{eqnarray}}

\DeclareRobustCommand{\SkipTocEntry}[4]{}

\newcommand{\nn}{\nonumber}

\newcommand\beal{\begin{aligned}}
\newcommand\eeal{\end{aligned}}

\usepackage{xcolor}
\newcommand\nl{\notag\\}

\newcommand{\bv}{{\boldsymbol v}}
\newcommand{\bSigma}{{\boldsymbol \Sigma}}
\newcommand{\bel}{{\boldsymbol \ell}}

\newcommand{\bk}{{\boldsymbol k}}

\newcommand{\bn}{{\boldsymbol n}}

\newcommand{\br}{{\boldsymbol r}}

\newcommand{\bx}{{\boldsymbol x}}

\newcommand{\bS}{{\boldsymbol S}}


\def\addDESY{Deutsches Elektronen-Synchrotron DESY, Notkestr. 85, 22607 Hamburg, Germany}

\begin{document}
\preprint{DESY-22-004}
\preprint{ET-0001A-22}

\title{Gravitational radiation from inspiralling compact objects:\\ [.2cm] Spin effects to fourth Post-Newtonian order}
\author{Gihyuk Cho}
\affiliation{\addDESY}
\author{Rafael A.\ Porto}
\affiliation{\addDESY}
\author{Zixin Yang}
\affiliation{\addDESY}

\begin{abstract}
The  linear-~and~quadratic-in-spin contributions to the binding potential and gravitational-wave flux from binary systems are derived to next-to-next-to-leading order in the Post-Newtonian (PN) expansion of general relativity, including finite-size and tail effects. The calculation is carried out through the worldline effective field theory framework. We find agreement in the overlap with the available PN literature and test-body limit. As a direct application, we complete the knowledge of spin effects in the evolution of the orbital phase for aligned-spin circular orbits to fourth PN~order. We estimate the impact of the new results in the number of accumulated gravitational-wave cycles. We~find they will play an important role in providing reliable physical interpretation of gravitational-wave signals from spinning binaries with future detectors such as LISA and the Einstein Telescope. 
\end{abstract}
\maketitle
\emph{Introduction.}  The dynamical evolution of compact binaries has been the main cause of the gravitational waves (GWs) detected by the LIGO-Virgo-KAGRA interferometers \cite{LIGOScientific:2021djp,Nitz:2021zwj,Olsen:2022pin}, and will continue to be one of the primary sources for future GW observatories such as the Laser Interferometer Space Antenna (LISA)  \cite{lisa} and the Einstein Telescope (ET) \cite{et}. The GWs produced by the inspiral, merger, and ring-down from the expected several two-body events will carry vast amounts of information that can shed light on long-standing problems in astrophysics, cosmology, and particle physics \cite{tune,music,Maggiore:2019uih,eucapt}.  In particular the spin of the constituents, which has been found to be large in several recent detections~\cite{Olsen:2022pin},  is not only strongly correlated with different formation channels, e.g. \cite{Kushnir:2016zee,Farr:2017uvj,Roulet:2021hcu,Callister:2021fpo,Bouffanais:2021wcr,Franciolini:2021xbq,Olsen:2022pin}, also offers a window to physics beyond the standard model, e.g. \cite{axiverse,gcollider1,gcollider2,salvo,salvo2,LIGOScientific:2021jlr,Ding:2020bnl,Takahashi:2021eso,Maggio:2021ans}. Therefore, high-precision waveforms incorporating spin corrections are an essential ingredient to exploit the discovery potential in GW~astronomy.\vskip 4pt After a concerted effort involving both numerical \cite{Ajith:2012az,Szilagyi:2015rwa,Dietrich:2018phi,Barack:2018yvs} as well as analytic techniques \cite{blanchet,Schafer:2018kuf,review}, GW template banks have been successfully used to analyse the GW data collected thus far \cite{LIGOScientific:2021djp,Nitz:2021zwj,Olsen:2022pin}. However, while current templates may be sufficient for detection, when it comes to parameter estimation, the formidable empirical reach of future experiments require higher levels of accuracy, both for the Post-Newtonian (PN) inspiral regime as well as merger stages of the binary's dynamics \cite{Buonanno:2009zt,Purrer:2019jcp,Ferguson:2020xnm,Galaudage:2021rkt}. Presently, although partial results for the derivation of the evolution of the orbital phase in the inspiral regime are known at 4PN order for non-spinning bodies through various computations \cite{Damour:2014jta,Jaranowski:2015lha,Bernard:2015njp, Bernard:2017bvn,Marchand:2017pir,damour3n,tail,apparent,lamb,nrgr4pn1,nrgr4pn2,Marchand:2020fpt,Henry:2021cek}, and even higher orders in the conservative sector \cite{Blumlein:2020pyo,Blumlein:2021txe,hered1,hered2,blum,blum2,binidam1,binidam2}, spin contributions have not been pushed so far to the same relative level of accuracy. In particular, for radiative effects, while spin-orbit corrections were obtained to next-to-next-to-leading order (N$^2$LO) \cite{nnloso1,nnloso2}, complete spin-spin effects are only known to NLO \cite{nrgrs,prl,nrgrs1,nrgrs2,srad,amps,bohenloss,Cho:2021mqw}. 
In this letter we fill this gap and report the completion of spin effects to N$^2$LO in the PN expansion and to quadratic order in the spins, corresponding to the 4PN order for rapidly rotating bodies.\vskip 4pt

The derivation involves several ingredients, which we obtain using the worldline effective field theory (EFT) framework in the PN regime \cite{nrgr,andirad} extended to spinning bodies \cite{nrgrs,prl,nrgrs1,nrgrs2,nrgrso,srad,amps}. The EFT approach uses powerful tools from particle physics resembling, for instance, methodologies used in the calculation of binding energies for heavy quark states \cite{IraTasi}. The problem of motion is thus reduced to a series of Feynman diagrams, involving potential and radiation modes, which are constructed by iteratively solving for the (classical) gravitational field sourced by compact objects treated as point-like objects. Utilizing the EFT formalism, we have compute the gravitational potential and necessary radiative multipole moments at linear and quadratic order in the spins entering in the flux to N$^2$LO, including finite-size effects. The~completion of spin contributions at 4PN entails also the so-called {\it tail effect}, due to the scattering of the outgoing radiation off of the background geometry, e.g. \cite{tail}, which we incorporate through the EFT approach. The values for all of the intermediate (very lengthy) results are displayed in the ancillary file, see also the supplemental material.  Perfect agreement is found in the overlap with previous results in the literature~\cite{nnloso1,nnloso2,tailuc,Levi:2016ofk,Antonelli:2020aeb,Antonelli:2020ybz,Khalil:2021fpm,pmefts,Kosmopoulos:2021zoq}. \vskip 4pt From the binding energy and radiated flux we derive the imprint of spin effects in the orbital phase evolution for aligned-spins circular orbits.~As a measure of the impact of the new terms, we estimate the accumulated GW cycles for various paradigmatic astrophysical configurations as well as detector-sensitivity curves. We~find the N$^2$LO spin terms make a significant contribution  both in ET and LISA frequency bands. (The effect increases the larger the mass ratio, which coincidentally is expected to correlate with larger spins~\cite{Callister:2021fpo}.) This is partially driven by quadratic-in-spin terms carrying information about the inner structure of the compact objects. The results presented here will therefore play an important role in elucidating the origin of binary black holes as well as aiding future discoveries, such as new dark objects \cite{Maggio:2021ans} or clouds of putative ultralight particles induced by superradiance \cite{axiverse,gcollider1,gcollider2,salvo,salvo2,LIGOScientific:2021jlr,Ding:2020bnl,Takahashi:2021eso}, through GW precision~data. \vskip 4pt

{\it Worldline EFT approach.} The effective action is obtained in the weak-field regime by \emph{solving for} the metric perturbation in the (classical) saddle-point approximation \cite{nrgr,nrgrs}. The compact bodies are described by the {\it Routhian}, 
\begin{align}
{\cal R} &= - \frac{1}{2}\sum_{n=1,2}   \bigg( m_n \, g_{\mu\nu} v_n^\mu v_n^\nu +  \omega_\mu^{ab}S_{n\, ab} v_n^{\mu}  \label{routhian}\\ &- \frac{C^{(n)}_{\rm ES^2}}{m_n} \frac{E_{ab} S_{n}^{ac}{S_{n\, c}}^b}{\sqrt{v_n^\mu v_{n\, \mu}}}  + \frac{1}{m_n} R_{de ab}S_n^{ab}S_n^{cd} v_n^e v_{n\, c}  +\cdots  \bigg)\,,\nn
\end{align}
which serves both as a Lagrangian for the position variables $(x_n^\alpha,v_n^\alpha)$ and a Hamiltonian for the spin, projected onto a locally-flat frame, $S_n^{ab} \equiv e^a_\mu e^b_\nu S_n^{\mu\nu}$, and coupled to $\omega_\mu^{ab}$, the Ricci rotation coefficients. The free parameters include the masses, $m_n$, as well as~$C^{(n)}_{\rm ES^2}$, which accounts for finite-size effects \cite{nrgrs,prl,nrgrs1,nrgrs2}. The latter couple to $E_{ab}$, the electric component of the Weyl tensor. The last term in \eqref{routhian}, involving the Riemann tensor, ensures the covariant spin-supplementarity-condition, $S^{ab} v_b=0$, is preserved upon evolution. However, for convenience, our results will be presented in terms of (Newton-Wigner) precession-only spin variables. The ellipses encapsulate higher orders in spin and curvature, which are not relevant in this letter. See \cite{review} for more~details. \vskip 4pt

\begin{figure}[t!] 
\includegraphics[width=0.27\textwidth]{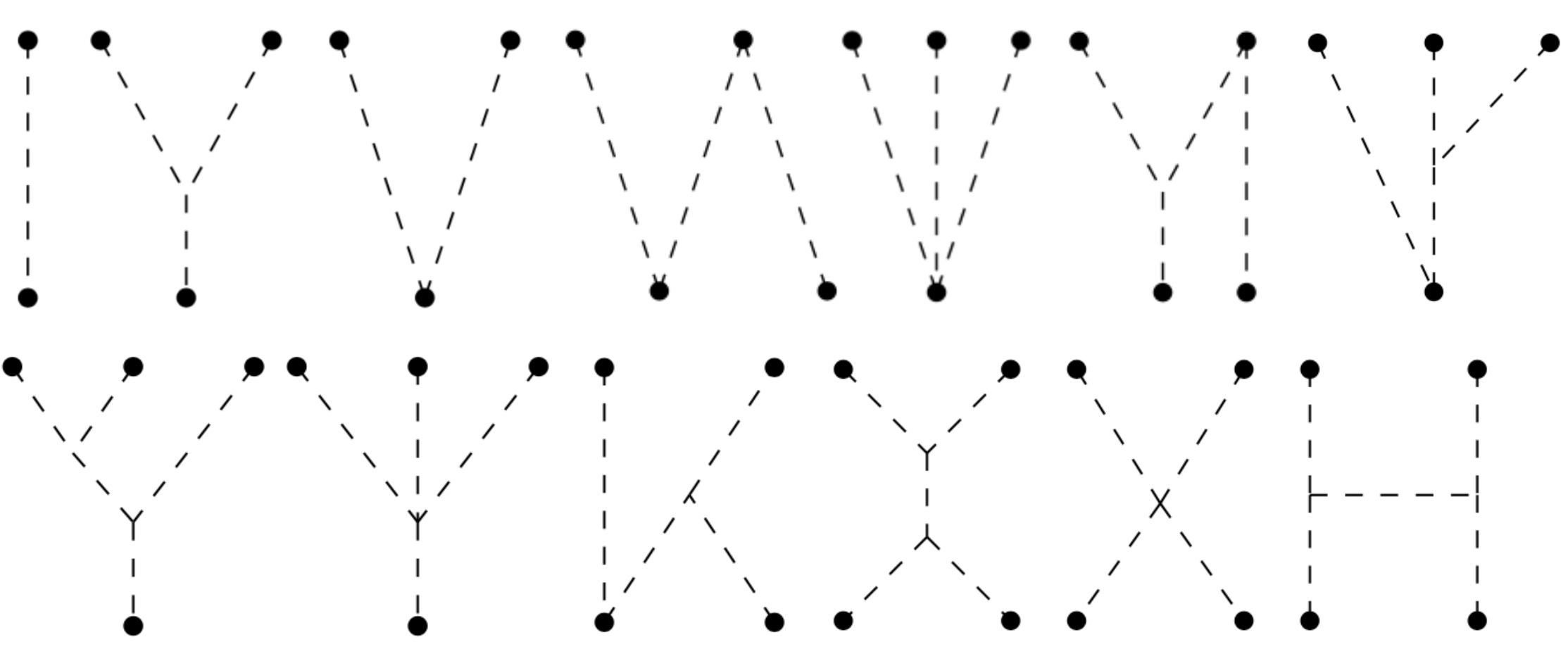} 
      \caption{Topologies needed for the N$^2$LO potential (see text). } 
      \label{fig1}
\end{figure}

{\it Gravitational potential \& Binding energy.} The derivation of the potential follows by computing the `vacuum-to-vacuum' amplitude in the presence of external sources, by {\it integrating out} the off-shell quasi-instantaneous modes of the gravitational field. The associated Feynman diagrams needed to N$^2$LO are depicted in Fig.~\ref{fig1}. The worldline couplings, depicted as a black disc, include the mass as well as the linear and quadratic spin terms shown in~\eqref{routhian}.  The dashed lines represent the potential modes of the gravitational field responsible for the binding of the two-body system. Hence, the Green's function (a.k.a. propagator) must be PN expanded \cite{nrgr}
\beq
\frac{i}{(k^0)^2-\bk^2} = -\frac{i}{\bk^2}\biggl(1+ \frac{(k^0)^2}{\bk^2}+\cdots\biggr)\,,\label{expot}
\eeq
with each factor of $(k^0)^2$ scaling as $v^2$. \vskip 4pt
From the gravitational potential, and after transforming to conserved-norm spins, we derive the equations of motion including the spin precession, which are displayed in full glory in the ancillary file. We then compute the binding energy, which can be written as 
\begin{align}
E_\text{SO}&= \frac{G M\nu}{r^2}\bigg[e_3^0 + \frac{1}{2} \left(e_5^0+\frac{G M}{r} e_5^1\right)\bigg]\nl
&+\frac{GM \nu}{4r^2}\,\bigg[\,\frac{1}{2}\left(e_7^0+\frac{G M}{r} e_7^1\right)+\frac{G^2M^2}{r^2}e_7^2\bigg]\,,\label{eso}
\end{align}
for the spin-orbit terms, and
\begin{align}
E_\text{SS}&=  \frac{G\nu}{4r^3}\,\bigg[e_4^0+\frac{1}{2}\,e_6^0+\frac{GM}{r} e_6^1\,\bigg]\,\nl
&+ \frac{G\nu}{16 r^3}\,\bigg[\,\frac{1}{2}\,e_8^0+\frac{GM}{r} e_8^1+\frac{7}{2}\,\frac{G^2M^2}{r^2}e_8^2\bigg]\,,\label{ess}
\end{align}
for spin-spin contribution. The value for the $e^i_j$  PN coefficients are given in the supplemental material and ancillary file. We use $M\equiv m_1+m_2$ for the total mass, $\nu \equiv m_1m_2/M^2$ for the symmetric-mass-ratio and $r\equiv |\bx_1-\bx_2|$ the relative distance, respectively. We use the parameter $\delta\equiv (m_1-m_2)/M$, and $\kappa_\pm \equiv C_{\rm ES^2}^{(1)} \pm C^{(2)}_{\rm ES^2}$. For the spin variables we use the standard, e.g. \cite{blanchet}, 
\beq
\begin{aligned}
\bS &\equiv \bS_1+\bS_2\,, \\
\bSigma &\equiv M\Big(\frac{\bS_2}{m_2}-\frac{\bS_1}{m_1}\Big)\,.\label{sigma} 
\end{aligned}
\eeq

\begin{figure}[t!] 
\includegraphics[width=0.25\textwidth]{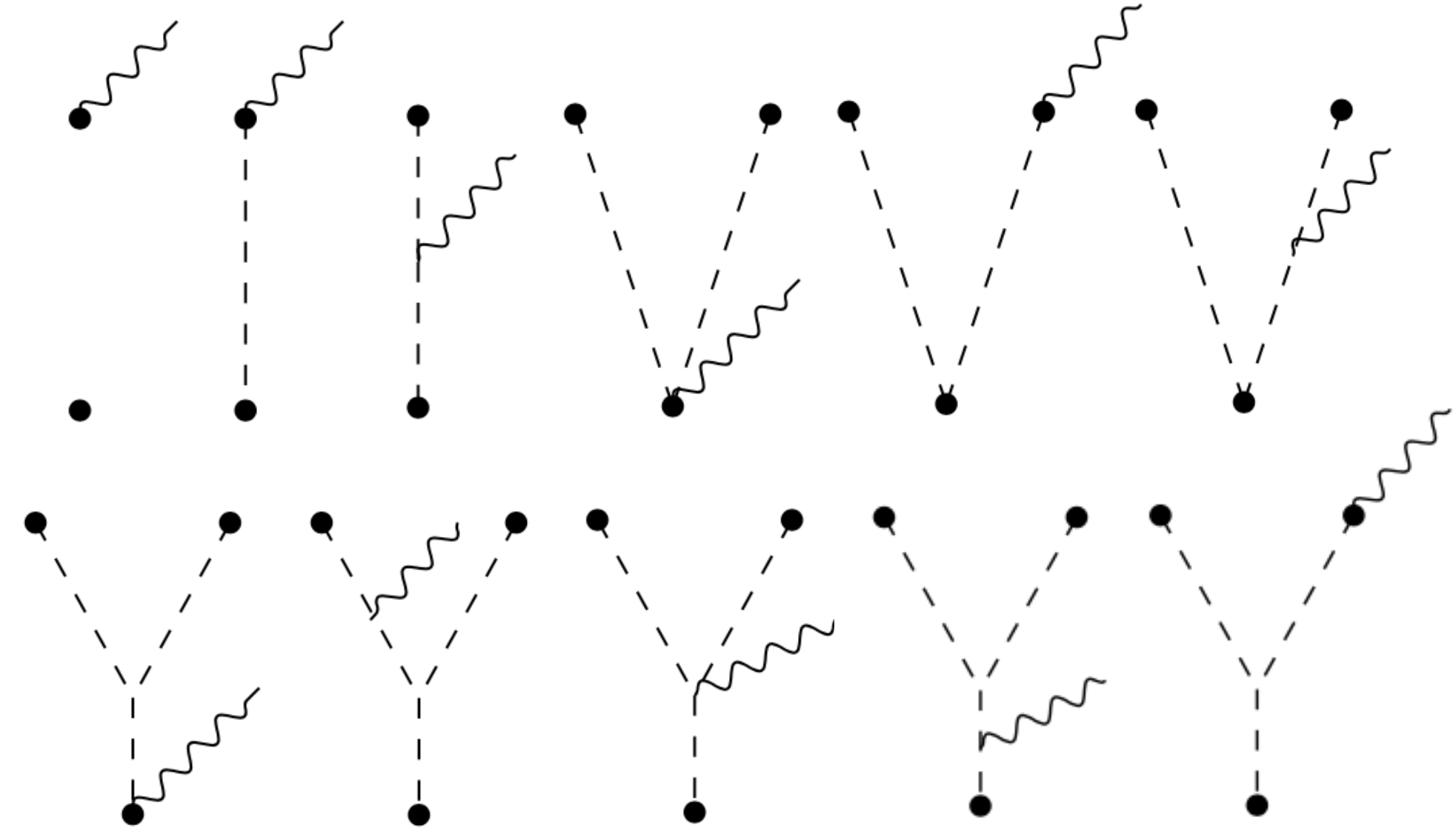} 
      \caption{Topologies needed to match the multipole moments entering the radiated flux to NNLO (see text).}
      \label{fig2}
\end{figure}

\begin{figure}[t!] 
\includegraphics[width=0.08\textwidth]{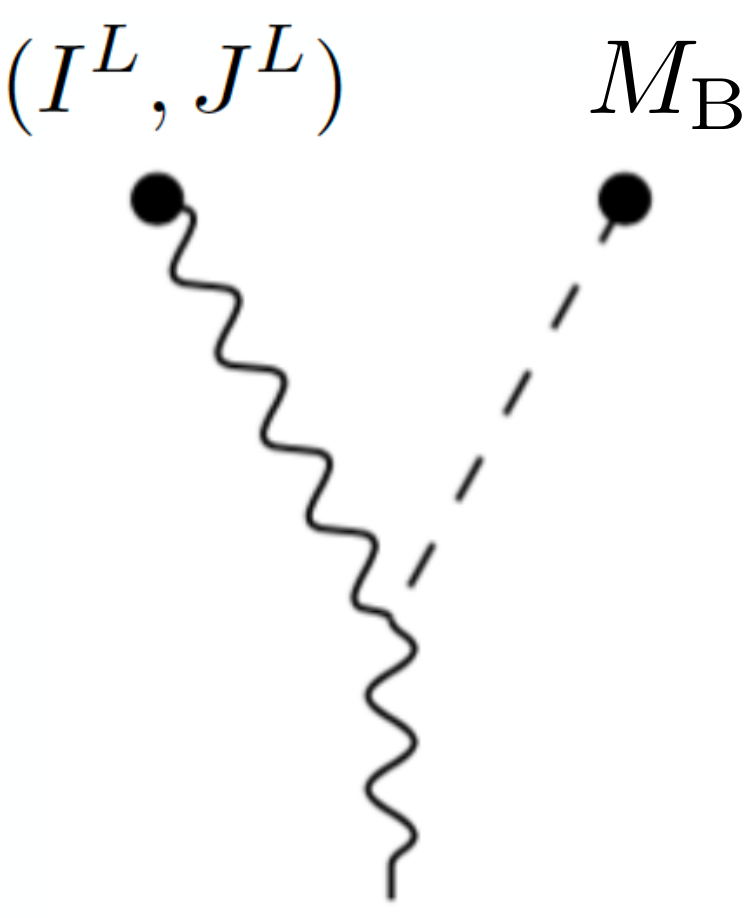} 
      \caption{Tail coupling between the binary's mass monopole and other source moments.}
      \label{fig3}
\end{figure}

{\it Radiated flux.}~The emitted power is obtained by matching the one-point function to a long-distance worldline effective theory for the binary system treated as a point-like object endowed with multipole moments. The~action includes, in addition to the (Bondi) mass-energy monopole term, $-M_{\rm B} \int d\tau$,  a series of symmetric-trace-free (STF) {\it source} mass, $I^L$, and current, $J^L$, time-dependent multipoles, with $L \equiv \{i_1\cdots i_L\}$, \cite{andirad}
\beq
\sum_{\ell=2} \biggl( \frac{1}{\ell!} I^L_{\rm STF} \nabla_{L-2} \bar E_{i_{\ell-1}i_\ell} - \frac{2\ell}{(2\ell+1)!}J_{\rm STF} ^L \nabla_{L-2} \bar B_{i_{\ell-1}i_\ell}\biggr)\biggr]\,,
\eeq 
which couple to (covariant) derivatives of $\bar E_{ij}$ and $\bar B_{kl}$, the electric and magnetic components of the Weyl tensor involving only the {\it background} radiation field. The relevant  topologies are shown in Fig.~\ref{fig2}. The wavy line represents the on-shell radiation, which couples both to the constituents of the binary as well as the binding modes. From the source multipoles we compute the energy flux by squaring the emission amplitude \cite{andirad}, 
\begin{align}
&{\cal F}_{\rm src} =G\,\biggl(\frac{1}{5}I^{(3)}_{ij}\,I^{(3)}_{ij}+\frac{16}{45}\,J^{(3)}_{ij}\,J^{(3)}_{ij}+\frac{1}{189}I^{(4)}_{ijk}I^{(4)}_{ijk}\label{flux1}\\
&+\frac{1}{84}J^{(4)}_{ijk}\,J^{(4)}_{ijk} +\frac{1}{9072}\,I^{(5)}_{ijkl}I^{(5)}_{ijkl}+\frac{4}{14175}\,J^{(5)}_{ijkl}J^{(5)}_{ijkl}\cdots\biggr)\,,\nn
\end{align}  
to the desired order. The time derivatives are computed within the adiabatic approximation, by using the conservative equations of motion.\vskip 4pt We find for the spin-orbit flux
\begin{align}
\label{fso}
&{\cal F}^\text{SO}_{\rm src}=\frac{8G^3M^3\nu^2}{15 r^5}\,\biggr(\bigg[ f_3^0+4 \frac{G M}{r} f_3^1\bigg]\\ &+\frac{1}{28}\bigg[ f_5^0+2\frac{G M}{r} f_5^1+ 4\frac{G^2M^2}{r^2} f_5^2\bigg]\nn \\
&+\frac{1}{84}\bigg[ f_7^0+\frac{GM}{r} f_7^1+\frac{2}{3}\frac{G^2M^2}{r^2} f_7^2+\frac{4}{9}\frac{G^3 M^3}{r^3} f_7^3\bigg]\biggr)\,,\notag
\end{align}
whereas, for the spin-spin terms, we have
\beq
\begin{aligned}
&{\cal F}^\text{SS}_{\rm src}=\frac{2G^3M^2\nu^2}{15r^6}\,\biggr(f_4^0+ \frac{1}{7}\bigg[f_6^0+\frac{GM}{r} f_6^1+ 4\frac{G^2M^2}{r^2} f_6^2\bigg]\\
&+ \frac{1}{84}\bigg[f_8^0+\frac{GM}{r}f_8^1+\frac{2}{3} \frac{G^2M^2}{r^2} f_8^2+\frac{8}{3}\frac{G^3 M^3}{r^3} f_8^3\bigg]\biggr)\,.\label{fss}
\end{aligned}
\eeq
The value of the $f^i_j$ coefficients are displayed in the supplemental material and ancillary file.\vskip 4pt  In order to complete the derivation of the total flux we must also include the tail effect, depicted in Fig.~\ref{fig3}, where the radiated field interacts with the background geometry around the binary system, sourced by the monopole term. This is often packaged in terms of  {\it radiative} multipole moments, which can then be used to compute the total power using \eqref{flux1}. The (leading) tail term yields
\begin{align}
&{\cal F}_\text{tail}=-G^2 M_{\rm B}\, \pi \int \bigg[\Big( \frac{2}{5}I_{ij}(p)\,I_{ij}(q) + \frac{32}{45}J_{ij}(p)\,J_{ij}(q) \Big)\nl
&-pq\Big( \frac{2}{189}I_{ijk}(p)\,I_{ijk}(q) + \frac{1}{42}J_{ijk}(p)\,J_{ijk}(q) \Big)\bigg]\times \nl &\quad\quad\quad p^3q^4\,\text{sign}(q)e^{-i(p+q)t}  dp \, dq\,,
\end{align}
where $I_L(p)$ is the Fourier transform.
In the above expression the $M_{\rm B}$ includes not only the total mass of the binary but also the kinetic energy and binding potential to a given PN order.\vskip 4pt

{\it Aligned-spin circular orbits.} As a direct application of our results we consider the phenomenologically relevant case of (planar) circular orbits with the (conserved-norm) spins being either aligned or anti-aligned with the angular momentum. In what follow we quote the results using the following projected (dimensionless) spin variables 
\begin{align}
\hat{S}_\ell\equiv \frac{{\bel}\cdot\bS}{G M^2}\,,\quad {\Sigma}_\ell\equiv\frac{{\bel}\cdot\bSigma}{G M^2}\,,
\end{align}
with $\bel$ the unit vector in the direction of the angular momentum. Garnering all the pieces together we find for the spin-dependent linear- and bilinear-in-spin binding energy as a function of the orbital frequency, $x\equiv (GM\Omega)^{2/3}$,  \begin{widetext}
\begingroup
\allowdisplaybreaks
\scriptsize
\begin{align}
&E_{\rm spin}=-\frac{M\,\nu\,x}{2}\Bigg[x^{3/2}\,\Big(\frac{14}{3} \,\hat{S}_\ell + 2 \,\delta \,\hat{\Sigma}_\ell\Big) + x^{2} \,\Bigg\{  \,\Big(-2 -  \,\kappa_{+}\Big) \,\hat{S}_\ell^2 
+ \,\bigg[\,\kappa_{-} + \,\delta \,\Big(-2 -  \,\kappa_{+}\Big)\bigg] \,\hat{S}_\ell \,\hat{\Sigma}_\ell \\ & + \,\bigg[2 \,\nu + \frac{1}{2} \,\delta \,\kappa_{-} + \,\Big(- \frac{1}{2} + \,\nu\Big) \,\kappa_{+}\bigg] \,\hat{\Sigma}_\ell^2\Bigg\}+x^{5/2} \,\bigg[\,\Big(11 -  \frac{61}{9} \,\nu\Big) \,\hat{S}_\ell + \,\Big(3 -  \frac{10}{3} \,\nu\Big) \,\delta \,\hat{\Sigma}_\ell\bigg] \nl
& + x^{3} \,\Bigg(\,\bigg[\frac{50}{9} + \frac{5}{3} \,\nu -  \frac{5}{3} \,\delta \,\kappa_{-} + \,\Big(- \frac{25}{6} + \frac{5}{6} \,\nu\Big) \,\kappa_{+}\bigg] \,\hat{S}_\ell^2
 + \,\Bigg\{\,\Big(\frac{5}{2} + \frac{35}{6} \,\nu\Big) \,\kappa_{-} + \,\delta \,\bigg[\frac{25}{3} + \frac{5}{3} \,\nu + \,\Big(- \frac{5}{2} + \frac{5}{6} \,\nu\Big) \,\kappa_{+}\bigg]\Bigg\} \,\hat{S}_\ell \,\hat{\Sigma}_\ell\nl
& + \,\bigg[5 - 10 \,\nu -  \frac{5}{3} \,\nu^2 + \,\Big(\frac{5}{4} + \frac{5}{4} \,\nu\Big) \,\delta \,\kappa_{-} + \,\Big(- \frac{5}{4} + \frac{5}{4} \,\nu -  \frac{5}{6} \,\nu^2\Big) \,\kappa_{+}\bigg] \,\hat{\Sigma}_\ell^2\Bigg) + x^{7/2}\,\bigg[\,\Big(\frac{135}{4} -  \frac{367}{4} \,\nu + \frac{29}{12} \,\nu^2\Big) \,\hat{S}_\ell + \,\Big(\frac{27}{4} - 39 \,\nu + \frac{5}{4} \,\nu^2\Big) \,\delta \,\hat{\Sigma}_\ell\bigg]\nl
&  + x^{4}\,\Bigg(\,\bigg[\frac{67}{12} + \frac{6445}{108} \,\nu -  \frac{7}{36} \,\nu^2 + \,\Big(- \frac{31}{4} + \frac{35}{18} \,\nu\Big) \,\delta \,\kappa_{-} + \,\Big(- \frac{125}{8} + \frac{1025}{72} \,\nu -  \frac{7}{72} \,\nu^2\Big) \,\kappa_{+}\bigg] \,\hat{S}_\ell^2 + \,\Bigg\{\,\Big(\frac{63}{8} + \frac{449}{24} \,\nu -  \frac{553}{72} \,\nu^2\Big) \,\kappa_{-} \nl
&+ \,\delta \,\bigg[\frac{49}{4} + \frac{1649}{36} \,\nu -  \frac{7}{36} \,\nu^2+ \,\Big(- \frac{63}{8} + \frac{295}{24} \,\nu -  \frac{7}{72} \,\nu^2\Big) \,\kappa_{+}\bigg]\Bigg\} \,\hat{S}_\ell \,\hat{\Sigma}_\ell + \,\bigg[\frac{21}{2} -  \frac{119}{12} \,\nu -  \frac{135}{4} \,\nu^2 + \frac{7}{36} \,\nu^3 + \,\Big(\frac{63}{16} + \frac{77}{48} \,\nu -  \frac{91}{48} \,\nu^2\Big) \,\delta \,\kappa_{-} \nl
&+ \,\Big(- \frac{63}{16} + \frac{301}{48} \,\nu -  \frac{499}{48} \,\nu^2 + \frac{7}{72} \,\nu^3\Big) \,\kappa_{+}\bigg] \,\hat{\Sigma}_\ell^2\Bigg)\Bigg]\,\notag.
\end{align}
\endgroup
On the other hand, the energy flux becomes
\begingroup
\allowdisplaybreaks
\scriptsize
\begin{align}
&{\cal F}_{\rm spin}=\frac{32\,\nu^2x^5}{5\,G}\,\Bigg[ x^{3/2} \,\Big(-4 \,\hat{S}_\ell -  \frac{5}{4} \,\delta \,\hat{\Sigma}_\ell\Big) + x^2\,\Bigg\{ \,\Big(4 + 2 \,\kappa_{+}\Big) \,\hat{S}_\ell^2 + \,\bigg[-2 \,\kappa_{-} + \,\delta \,\Big(4 + 2 \,\kappa_{+}\Big)\bigg] \,\hat{S}_\ell \,\hat{\Sigma}_\ell \\ & + \,\bigg[\frac{1}{16} - 4 \,\nu-  \,\delta \,\kappa_{-} + \,\kappa_{+} \,\Big(1 - 2 \,\nu\Big) \bigg] \,\hat{\Sigma}_\ell^2\Bigg\}
 + x^{5/2} \,\bigg[\,\Big(- \frac{9}{2} + \frac{272}{9} \,\nu\Big) \,\hat{S}_\ell + \,\delta \,\Big(- \frac{13}{16} + \frac{43}{4} \,\nu\Big) \,\hat{\Sigma}_\ell\bigg] \nl
& +  x^3\,\Bigg(-16 \,\pi \,\hat{S}_\ell -  \frac{31}{6} \,\pi \,\delta \,\hat{\Sigma}_\ell+\,\bigg[- \frac{5239}{504} -  \frac{43}{2} \,\nu+ \frac{41}{16} \,\delta \,\kappa_{-} + \,\kappa_{+} \,\Big(- \frac{271}{112} -  \frac{43}{4} \,\nu\Big) \bigg] \,\hat{S}_\ell^2\nl
& + \,\Bigg\{\,\delta \,\bigg[- \frac{817}{56} + \,\kappa_{+} \,\Big(- \frac{279}{56} -  \frac{43}{4} \,\nu\Big) -  \frac{43}{2} \,\nu\bigg] + \,\kappa_{-} \,\Big(\frac{279}{56} + \frac{1}{2} \,\nu\Big)\Bigg\} \,\hat{S}_\ell \,\hat{\Sigma}_\ell \nl
&+ \,\bigg[- \frac{25}{8} + \frac{344}{21} \,\nu + \frac{43}{2} \,\nu^2 + \,\delta \,\kappa_{-} \,\Big(\frac{279}{112} + \frac{45}{16} \,\nu\Big) + \,\kappa_{+} \,\Big(- \frac{279}{112} + \frac{243}{112} \,\nu + \frac{43}{4} \,\nu^2\Big)\bigg] \,\hat{\Sigma}_\ell^2\Bigg) \nl
&+ x^{7/2}\,\Bigg\{\,\Big(\frac{476645}{6804} + \frac{6172}{189} \,\nu -  \frac{2810}{27} \,\nu^2\Big) \,\hat{S}_\ell + \,\delta \,\Big(\frac{9535}{336} + \frac{1849}{126} \,\nu -  \frac{1501}{36} \,\nu^2\Big) \,\hat{\Sigma}_\ell+\Big( 16\,\pi+8 \,\pi \,\kappa_{+}\Big) \,\hat{S}_\ell^2 \nl &+ \,\bigg[-8 \,\pi \,\kappa_{-} + \,\delta \,\Big(16 \,\pi + 8 \,\pi \,\kappa_{+}\Big)\bigg] \,\hat{S}_\ell \,\hat{\Sigma}_\ell + \,\bigg[\frac{1}{8} \,\pi - 4 \,\pi \,\delta \,\kappa_{-} - 16 \,\pi \,\nu + \,\kappa_{+} \,\Big(4 \,\pi - 8 \,\pi \,\nu\Big)\bigg] \,\hat{\Sigma}_\ell^2\Bigg\}\nl
&+  x^4 \,\Bigg(\Big(-\frac{3485\,\pi}{96} + \frac{13879\,\pi}{72} \,\nu \Big) \,\hat{S}_\ell + \,\delta \,\Big(-\frac{7163\,\pi}{672} + \frac{130583\,\pi}{2016} \,\nu\Big) \,\hat{\Sigma}_\ell
+\,\bigg[- \frac{4289}{648} -  \frac{295}{21} \,\nu + 54 \,\nu^2+ \,\delta \,\kappa_{-} \,\Big(\frac{935}{336} -  \frac{2153}{144} \,\nu\Big)  \nl
&+ \,\kappa_{+} \,\Big(- \frac{124577}{9072} + \frac{3265}{126} \,\nu + 27 \,\nu^2\Big)\bigg] \,\hat{S}_\ell^2+ \,\Bigg\{\,\kappa_{-} \,\Big(\frac{74911}{4536} -  \frac{52411}{1008} \,\nu + \frac{1181}{36} \,\nu^2\Big)\nl
& + \,\delta \,\bigg[- \frac{160621}{9072} + \frac{9977}{252} \,\nu + 54 \,\nu^2 + \,\kappa_{+} \,\Big(- \frac{74911}{4536} + \frac{41191}{1008} \,\nu + 27 \,\nu^2\Big)\bigg]\Bigg\} \,\hat{S}_\ell \,\hat{\Sigma}_\ell \nl
&+ \,\bigg[\frac{1633}{336} + \frac{465071}{18144} \,\nu -  \frac{74033}{1008} \,\nu^2 - 54 \,\nu^3 + \,\delta \,\kappa_{-} \,\Big(\frac{74911}{9072} -  \frac{46801}{2016} \,\nu + \frac{209}{144} \,\nu^2\Big) + \,\kappa_{+} \,\Big(- \frac{74911}{9072} + \frac{102979}{2592} \,\nu -  \frac{7109}{168} \,\nu^2 - 27 \,\nu^3\Big)\bigg] \,\hat{\Sigma}_\ell^2\Bigg)\Bigg]\,.\notag
\end{align}
\endgroup 
\end{widetext}
As a consistency check, this result agrees with the GW flux computed for a (non-spinning) test body orbiting around a Kerr black hole in \cite{test}, to the given PN order. We~then combine these results to derive the evolution of the orbital frequency, from which we infer the change in the orbital phase, $\phi = \int \Omega(t) dt $, using the TaylorT2 approximant, e.g. \cite{Buonanno:2009zt}, yielding
\begin{widetext}
\begingroup
\allowdisplaybreaks
\scriptsize
\begin{align}
&\phi_{\rm spin}=-\frac{x^{-5/2}}{32 \,\nu } \Bigg[\,x^{3/2} \,\Big(\frac{235}{6} \,\hat{S}_\ell + \frac{125}{8} \,\delta \,\hat{\Sigma}_\ell\Big)+ \,x^2 \,\Bigg\{ \,\Big(-50 - 25 \,\kappa_{+}\Big) \,\hat{S}_\ell^2 \\
&+ \,\bigg[25 \,\kappa_{-} + \,\delta \,\Big(-50 - 25 \,\kappa_{+}\Big)\bigg] \,\hat{S}_\ell \,\hat{\Sigma}_\ell + \,\bigg[- \frac{5}{16} + 50 \,\nu + \frac{25}{2} \,\delta \,\kappa_{-} + \,\Big(- \frac{25}{2} + 25 \,\nu\Big) \,\kappa_{+}\bigg] \,\hat{\Sigma}_\ell^2\Bigg\} \nl
& + \,x^{5/2} \log x \,\bigg[\,\Big(- \frac{554345}{2016} -  \frac{55}{8} \,\nu\Big) \,\hat{S}_\ell + \,\Big(- \frac{41745}{448} + \frac{15}{8} \,\nu\Big) \,\delta \,\hat{\Sigma}_\ell\bigg] \nl
&+ \,x^3 \,\Bigg(\frac{940}{3} \,\pi \,\hat{S}_\ell  + \frac{745}{6} \,\pi \,\delta \,\hat{\Sigma}_\ell + \,\bigg[- \frac{31075}{126} + 60 \,\nu + \frac{2215}{48} \,\delta \,\kappa_{-} + \,\Big(\frac{15635}{84} + 30 \,\nu\Big) \,\kappa_{+}\bigg] \,\hat{S}_\ell^2\nl
&+ \,\Bigg\{\,\Big(- \frac{47035}{336} -  \frac{2575}{12} \,\nu\Big) \,\kappa_{-} + \,\delta \,\bigg[- \frac{9775}{42} + 60 \,\nu + \,\Big(\frac{47035}{336} + 30 \,\nu\Big) \,\kappa_{+}\bigg]\Bigg\} \,\hat{S}_\ell \,\hat{\Sigma}_\ell\nl
& + \,\bigg[- \frac{410825}{2688} + \frac{23535}{112} \,\nu - 60 \,\nu^2 + \,\Big(- \frac{47035}{672} -  \frac{2935}{48} \,\nu\Big) \,\delta \,\kappa_{-} + \,\Big(\frac{47035}{672} -  \frac{4415}{56} \,\nu - 30 \,\nu^2\Big) \,\kappa_{+}\bigg] \,\hat{\Sigma}_\ell^2\Bigg) \nl
&+ \,x^{7/2} \,\Bigg\{\,\Big(- \frac{8980424995}{6096384} + \frac{6586595}{6048} \,\nu -  \frac{305}{288} \,\nu^2\Big) \,\hat{S}_\ell+ \,\Big(- \frac{170978035}{387072} + \frac{2876425}{5376} \,\nu + \frac{4735}{1152} \,\nu^2\Big) \,\delta \,\hat{\Sigma}_\ell \nl
&+ \,\Big(-100 \,\pi - 50 \,\pi \,\kappa_{+}\Big) \,\hat{S}_\ell^2  + \,\bigg[50 \,\pi \,\kappa_{-} + \,\delta \,\Big(-100 \,\pi - 50 \,\pi \,\kappa_{+}\Big)\bigg] \,\hat{S}_\ell \,\hat{\Sigma}_\ell + \,\bigg[- \frac{15}{16} \,\pi + 100 \,\nu \,\pi + 25 \,\pi \,\delta \,\kappa_{-} + \,\Big(-25 \,\pi + 50 \,\nu \,\pi\Big) \,\kappa_{+}\bigg] \,\hat{\Sigma}_\ell^2\Bigg\}\nl
&+ \,x^4 \,\Bigg(\Big(\frac{2388425\,\pi}{3024} - \frac{9925\,\pi}{36} \,\nu \Big) \,\hat{S}_\ell + \,\delta \,\Big(\frac{3237995\,\pi}{12096} - \frac{258245\,\pi}{2016} \,\nu\Big) \,\hat{\Sigma}_\ell +\,\bigg[- \frac{83427805}{72576} -  \frac{19720}{63} \,\nu + \frac{475}{24} \,\nu^2 + \,\Big(\frac{3284125}{24192} + \frac{1115}{72} \,\nu\Big) \,\delta \,\kappa_{-}\nl
& + \,\Big(\frac{55124675}{145152} -  \frac{32825}{756} \,\nu + \frac{475}{48} \,\nu^2\Big) \,\kappa_{+}\bigg] \,\hat{S}_\ell^2 + \,\Bigg\{\,\Big(- \frac{35419925}{145152} -  \frac{975955}{2016} \,\nu -  \frac{10345}{144} \,\nu^2\Big) \,\kappa_{-} \nl
&+ \,\delta \,\bigg[- \frac{66536845}{72576} -  \frac{109535}{378} \,\nu + \frac{475}{24} \,\nu^2 + \,\Big(\frac{35419925}{145152} -  \frac{89065}{1512} \,\nu + \frac{475}{48} \,\nu^2\Big) \,\kappa_{+}\bigg]\Bigg\} \,\hat{S}_\ell \,\hat{\Sigma}_\ell\nl
& + \,\bigg[- \frac{17815050265}{48771072} + \frac{26426305}{41472} \,\nu + \frac{12570535}{48384} \,\nu^2 -  \frac{475}{24} \,\nu^3 + \,\Big(- \frac{35419925}{290304} -  \frac{2571605}{24192} \,\nu -  \frac{5885}{288} \,\nu^2\Big) \,\delta \,\kappa_{-} \nl
& + \,\Big(\frac{35419925}{290304} -  \frac{19990295}{145152} \,\nu + \frac{479845}{6048} \,\nu^2 -  \frac{475}{48} \,\nu^3\Big) \,\kappa_{+}\bigg] \,\hat{\Sigma}_\ell^2\Bigg) \Bigg]\,,\notag
\end{align}
\endgroup
\end{widetext}
included in the ancillary file for the reader's convenience. We find perfect agreement in the overlap at linear order in the spin to 3.5PN of \cite{nnloso2,tailuc}, whereas spin-spin effects including finite-size corrections, both at 3.5PN and 4PN orders, are reported here for the first time. \vskip 4pt

{\it Conclusions and Outlook.} We have completed the knowledge of spin effects in the orbital phase evolution of compact binary systems to N$^2$LO in the PN expansion of general relativity and quadratic order in the spins, corresponding to an overall 4PN order for rapidly rotating bodies, including both finite-size as well as tail effects. The various ingredients for the full derivation, such as the gravitational potential and multipole moments, were obtained through the worldline EFT framework for spinning compact objects \cite{review}, which systematizes the two-body problem into a series of Feynman diagrams involving potential and radiation modes. Agreement is found in the overlap with previous PN results in the conservative \cite{Levi:2016ofk,Antonelli:2020aeb,Antonelli:2020ybz, Khalil:2021fpm,pmefts,Kosmopoulos:2021zoq} and radiation sectors \cite{nnloso1,nnloso2,tailuc}. In order to evaluate the impact of the new spin-dependent terms in the GW phase evolution, we used the leading quadrupolar approximation ($\phi_{\rm GW} \simeq 2 \phi$) to estimate the number of GW cycles in future detector's bands operating at design-sensitivity. The results, particularized to LISA (0.1mHz to {\it min}($f_\text{ISCO}$, 1 Hz))~and~ET (1Hz to $f_\text{ISCO}$, with $f_\text{ISCO}=\frac{1}{6^{3/2}\pi G M}$), are summarized here:\footnote{We have not included the known absorption \cite{poisson,dis2,Goldberger:2020fot}, radiation-reaction \cite{natalia1,natalia2} or cubic-in-spin \cite{Marsat:2014xea} effects.} 
\begin{widetext}
\begin{center}
\begin{figure}[h!] 
\vskip -0.2cm
\includegraphics[width=0.7\textwidth]{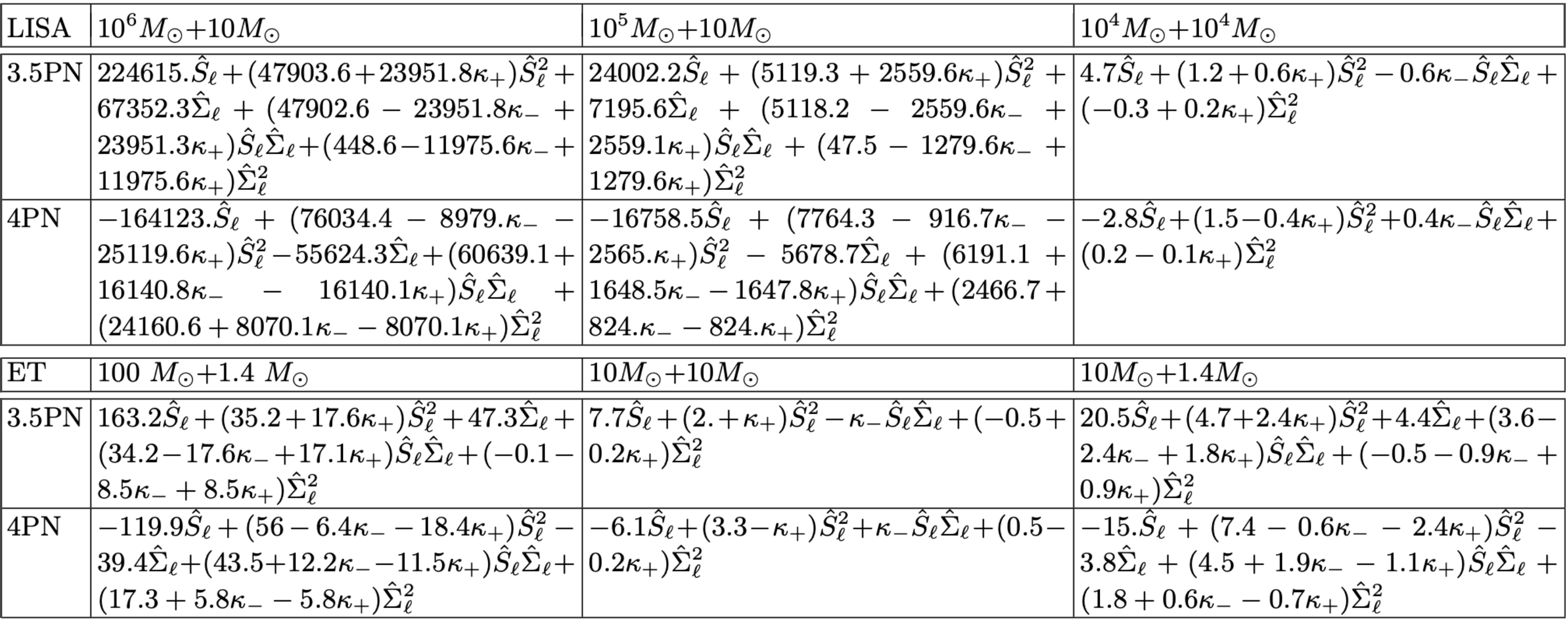} 
  \end{figure} 
\end{center}
\end{widetext}
Although generically relevant for comparable masses, spin effects become more important for binaries with unequal masses, which are also expected to exhibit larger spins \cite{Callister:2021fpo}. Moreover, terms quadratic in the spins, depending on the inner structure of compact bodies and nearby surroundings through the $\kappa_\pm$ couplings \cite{gcollider1,gcollider2}, account for a large portion of the accumulated GW~cycles. We therefore expect N$^2$LO spin terms to play an important role in providing accurate waveform models and reliable physical interpretation of GW signals from rotating compact binaries, thus motivating further in-depth studies to fully characterize their impact in detection and parameter estimation with forthcoming third-generation GW experiments.\vskip 4pt  

{\it Acknowledgments.} We are grateful to Brian Pardo for collaboration in \cite{Cho:2021mqw} and during the early stages of this project. We thank Gregor K\"alin for his help drawing the diagrams. This~work was supported by the ERC Consolidator Grant ``Precision Gravity: From the LHC~to LISA,"  provided by the European Research Council (ERC) under the European Union's H2020 research and innovation programme, grant No.\,817791. We~acknowledge extensive use of the \texttt{xAct} packages~\cite{martin2019xact}.
\newpage
\appendix
\section{Supplemental Material}
In what follows we summarize various results quoted in the main text. All these values are also conveniently given in the ancillary file. We use the (Euclidean) abbreviation
\beq
(ab) \equiv a^i b^i\,,\quad\quad (a,b,c)\equiv \epsilon^{ijk} a^i b^j c^k\,, 
\eeq
to simplify the notation. We also define the norm vector as $\bn \equiv \br/r$ and use $\bv \equiv \bv_1-\bv_2$ for the relative velocity. The coefficients of the spin-orbit and spin-spin binding energy in \eqref{eso}-\eqref{ess} are give by: 
\begin{widetext}
\begingroup
\allowdisplaybreaks
\scriptsize
\begin{align}
e_3^0&=- \,(n, S, v) -  \,\delta \,(n, \Sigma, v)\,,\nl
e_5^0&=3 \,\nu \,(nv)^2 \,(n, S, v) + 3 \,\big(1 + \,\nu\big) \,(n, S, v) \,v^2 + \,\big(-1 + 5 \,\nu\big) \,\delta \,(n, \Sigma, v) \,v^2\,,\nl
e_5^1&=-4 \,\nu \,(n, S, v) - 3 \,\nu \,\delta \,(n, \Sigma, v)\,,\nl
e_7^0&=15 \,\nu \,\big(-1 + 3 \,\nu\big) \,(nv)^4 \,(n, S, v) + 15 \,\nu^2 \,\delta \,(nv)^4 \,(n, \Sigma, v) + 6 \,\big(1 - 17 \,\nu\big) \,\nu \,(nv)^2 \,(n, S, v) \,v^2\nl
& - 12 \,\nu \,\big(2 + \,\nu\big) \,\delta \,(nv)^2 \,(n, \Sigma, v) \,v^2 + \,\big(21 - 31 \,\nu - 55 \,\nu^2\big) \,(n, S, v) \,v^4\nl
& + \,\big(-3 + 53 \,\nu - 99 \,\nu^2\big) \,\delta \,(n, \Sigma, v) \,v^4\nl
e_7^1&=2 \,\nu \,\big(187 + 20 \,\nu\big) \,(nv)^2 \,(n, S, v) + \,\nu \,\big(179 + 12 \,\nu\big) \,\delta \,(nv)^2 \,(n, \Sigma, v) \nl
&+ 2 \,\big(24 - 143 \,\nu + 40 \,\nu^2\big) \,(n, S, v) \,v^2 + \,\big(-16 - 145 \,\nu + 82 \,\nu^2\big) \,\delta \,(n, \Sigma, v) \,v^2\,,\nl
e_7^2&=\,\big(6 - 111 \,\nu - 8 \,\nu^2\big) \,(n, S, v) - 2 \,\big(-3 + 25 \,\nu + 3 \,\nu^2\big) \,\delta \,(n, \Sigma, v)\,,\nl
e_4^0&=6 \,\big(2 + \,\kappa_{+}\big) \,(nS)^2 + \,\big[-6 \,\kappa_{-} + 6 \,\delta \,\big(2 + \,\kappa_{+}\big)\big] \,(nS) \,(n\Sigma) \nl
&+ \,\big[-3 \,\delta \,\kappa_{-} + 3 \,\kappa_{+} - 6 \,\nu \,\big(2 + \,\kappa_{+}\big)\big] \,(n\Sigma)^2 - 2 \,\big(2 + \,\kappa_{+}\big) \,\bS^2 \nl
&+ \,\big[2 \,\kappa_{-} - 2 \,\delta \,\big(2 + \,\kappa_{+}\big)\big] \,(S\Sigma) + \,\big[\,\delta \,\kappa_{-} -  \,\kappa_{+} + 2 \,\nu \,\big(2 + \,\kappa_{+}\big)\big] \,\bSigma^2\,,\nl
e_6^0&=\,\big(28 + 6 \,\delta \,\kappa_{-} - 6 \,\kappa_{+}\big) \,(vS)^2 + 4 \,\big[\,\big(3 - 6 \,\nu\big) \,\kappa_{-} + \,\delta \,\big(13 - 3 \,\kappa_{+}\big)\big] \,(vS) \,(v\Sigma)\nl
& + \,\big[-2 \,\nu \,\big(38 + 3 \,\delta \,\kappa_{-} - 9 \,\kappa_{+}\big) + 6 \,\big(4 + \,\delta \,\kappa_{-} -  \,\kappa_{+}\big)\big] \,(v\Sigma)^2 \nl
&+ 6 \,\big[-14 - 3 \,\delta \,\kappa_{-} + 3 \,\kappa_{+} + 2 \,\nu \,\big(2 + \,\kappa_{+}\big)\big] \,(nv) \,(vS) \,(nS) + 6 \,\Big\{\,\big(-3 + 5 \,\nu\big) \,\kappa_{-} \nl
&+ \,\delta \,\big[-14 + 3 \,\kappa_{+} + \,\nu \,\big(2 + \,\kappa_{+}\big)\big]\Big\} \,(nv) \,(v\Sigma) \,(nS) + \,\Big\{-30 \,\nu \,\big(2 + \,\kappa_{+}\big) \,(nv)^2\nl
& - 6 \,\big[-10 + 3 \,\kappa_{+} + \,\nu \,\big(2 + \,\kappa_{+}\big)\big] \,v^2\Big\} \,(nS)^2 + 6 \,\big[\,\big(-3 + 5 \,\nu\big) \,\kappa_{-} \nl
&+ 3 \,\delta \,\big(-4 + \,\kappa_{+}\big) + \,\nu \,\delta \,\big(2 + \,\kappa_{+}\big)\big] \,(nv) \,(vS) \,(n\Sigma) \nl
&- 6 \,\big[-2 \,\nu \,\big(19 + \,\delta \,\kappa_{-} - 4 \,\kappa_{+}\big) + 3 \,\big(4 + \,\delta \,\kappa_{-} -  \,\kappa_{+}\big) + 2 \,\nu^2 \,\big(2 + \,\kappa_{+}\big)\big] \,(nv) \,(v\Sigma) \,(n\Sigma)\nl
& + \,\big\{30 \,\nu \,\big[\,\kappa_{-} -  \,\delta \,\big(2 + \,\kappa_{+}\big)\big] \,(nv)^2 - 6 \,\big[- \,\big(3 + \,\nu\big) \,\kappa_{-} + 3 \,\delta \,\big(-6 + \,\kappa_{+}\big)\nl
& + \,\nu \,\delta \,\big(2 + \,\kappa_{+}\big)\big] \,v^2\Big\} \,(nS) \,(n\Sigma) + \,\big\{15 \,\nu \,\big[\,\delta \,\kappa_{-} -  \,\kappa_{+} + 2 \,\nu \,\big(2 + \,\kappa_{+}\big)\big] \,(nv)^2 \nl
&+ \,\big[48 + 9 \,\delta \,\kappa_{-} - 9 \,\kappa_{+} + 6 \,\nu^2 \,\big(2 + \,\kappa_{+}\big) + 3 \,\nu \,\big(-52 + \,\delta \,\kappa_{-} + 5 \,\kappa_{+}\big)\big] \,v^2\Big\} \,(n\Sigma)^2\nl
& + \,\big\{6 \,\big[4 + \,\delta \,\kappa_{-} -  \,\kappa_{+} + \,\nu \,\big(2 + \,\kappa_{+}\big)\big] \,(nv)^2 + \,\big[-28 - 2 \,\delta \,\kappa_{-} + 8 \,\kappa_{+} + 2 \,\nu \,\big(2 + \,\kappa_{+}\big)\big] \,v^2\Big\} \,\bS^2 \nl
&+ \,\Big(6 \,\Big\{\,\big(2 - 5 \,\nu\big) \,\kappa_{-} + \,\delta \,\big[8 - 2 \,\kappa_{+} + \,\nu \,\big(2 + \,\kappa_{+}\big)\big]\Big\} \,(nv)^2 + 2 \,\Big\{\,\big(-5 + 3 \,\nu\big) \,\kappa_{-} \nl
&+ \,\delta \,\big[-26 + 5 \,\kappa_{+} + \,\nu \,\big(2 + \,\kappa_{+}\big)\big]\Big\} \,v^2\Big) \,(S\Sigma) + \,\Big\{-3 \,\big[-8 - 2 \,\delta \,\kappa_{-}\nl
& + \,\nu \,\big(24 + 3 \,\delta \,\kappa_{-} - 7 \,\kappa_{+}\big) + 2 \,\kappa_{+} + 2 \,\nu^2 \,\big(2 + \,\kappa_{+}\big)\big] \,(nv)^2 \nl
&+ \,\big[-24 - 5 \,\delta \,\kappa_{-} + \,\nu \,\big(76 + \,\delta \,\kappa_{-} - 11 \,\kappa_{+}\big) + 5 \,\kappa_{+} - 2 \,\nu^2 \,\big(2 + \,\kappa_{+}\big)\big] \,v^2\Big\} \,\bSigma^2\,,\nl
e_6^1&=\,\big(-36 + 9 \,\delta \,\kappa_{-} - 15 \,\kappa_{+}\big) \,(nS)^2 - 4 \,\big(8 \,\delta - 6 \,\kappa_{-} + 9 \,\nu \,\kappa_{-} + 6 \,\delta \,\kappa_{+}\big) \,(nS) \,(n\Sigma) \nl
&+ \,\big[12 \,\big(\,\delta \,\kappa_{-} -  \,\kappa_{+}\big) + \,\nu \,\big(30 - 9 \,\delta \,\kappa_{-} + 33 \,\kappa_{+}\big)\big] \,(n\Sigma)^2 + \,\big(8 - 3 \,\delta \,\kappa_{-} + 5 \,\kappa_{+}\big) \,\bS^2 \nl
&+ \,\big[4 \,\big(-2 + 3 \,\nu\big) \,\kappa_{-} + 8 \,\delta \,\big(1 + \,\kappa_{+}\big)\big] \,(S\Sigma)\nl
& + \,\big[-4 \,\delta \,\kappa_{-} + \,\nu \,\big(-10 + 3 \,\delta \,\kappa_{-} - 11 \,\kappa_{+}\big) + 4 \,\kappa_{+}\big] \,\bSigma^2\,,\nl
e_8^0&=\,\big\{-6 \,\big[-8 - 3 \,\delta \,\kappa_{-} + 3 \,\kappa_{+} + 6 \,\nu^2 \,\big(2 + \,\kappa_{+}\big) - 2 \,\nu \,\big(-26 + 6 \,\delta \,\kappa_{-} + 7 \,\kappa_{+}\big)\big] \,(nv)^2\nl
& - 4 \,\big[-39 - 9 \,\delta \,\kappa_{-} + \,\nu \,\big(70 + 27 \,\delta \,\kappa_{-} - 15 \,\kappa_{+}\big) + 9 \,\kappa_{+}\big] \,v^2\big\} \,(vS)^2\nl
& + \,\big\{6 \,\big[8 + 3 \,\delta \,\kappa_{-} - 2 \,\nu \,\big(12 + 2 \,\delta \,\kappa_{-} - 5 \,\kappa_{+}\big) - 3 \,\kappa_{+}\nl
& + 6 \,\nu^3 \,\big(2 + \,\kappa_{+}\big) + \,\nu^2 \,\big(-44 - 9 \,\delta \,\kappa_{-} + 7 \,\kappa_{+}\big)\big] \,(nv)^2 + 4 \,\big[\,\nu^2 \,\big(286 + 27 \,\delta \,\kappa_{-} - 69 \,\kappa_{+}\big) \nl
&- 3 \,\nu \,\big(65 + 10 \,\delta \,\kappa_{-} - 16 \,\kappa_{+}\big) + 9 \,\big(4 + \,\delta \,\kappa_{-} -  \,\kappa_{+}\big)\big] \,v^2\big\} \,(v\Sigma)^2\nl
& + \,\big\{60 \,\nu \,\big[10 - 3 \,\delta \,\kappa_{-} - 5 \,\kappa_{+} + 6 \,\nu \,\big(2 + \,\kappa_{+}\big)\big] \,(nv)^3\nl
& - 6 \,\big[86 + 21 \,\delta \,\kappa_{-} - 2 \,\nu \,\big(90 + 30 \,\delta \,\kappa_{-} - 11 \,\kappa_{+}\big) - 21 \,\kappa_{+} + 54 \,\nu^2 \,\big(2 + \,\kappa_{+}\big)\big] \,(nv) \,v^2\big\} \,(vS) \,(nS)\nl
& + \,\Big(60 \,\nu \,\big[\,\kappa_{-} + 3 \,\nu \,\kappa_{-} + \,\big(-1 + 3 \,\nu\big) \,\delta \,\big(2 + \,\kappa_{+}\big)\big] \,(nv)^3\nl
& - 6 \,\big\{\,\big(21 - 83 \,\nu + 93 \,\nu^2\big) \,\kappa_{-} + \,\delta \,\big[86 - 21 \,\kappa_{+} + 27 \,\nu^2 \,\big(2 + \,\kappa_{+}\big) \nl
&+ \,\nu \,\big(-214 + 41 \,\kappa_{+}\big)\big]\big\} \,(nv) \,v^2\Big) \,(v\Sigma) \,(nS) + \,\big\{-210 \,\nu \,\big(-1 + 3 \,\nu\big) \,\big(2 + \,\kappa_{+}\big) \,(nv)^4\nl
& + 60 \,\nu \,\big[-18 -  \,\kappa_{+} + 15 \,\nu \,\big(2 + \,\kappa_{+}\big)\big] \,(nv)^2 \,v^2 + 6 \,\big[54 - 21 \,\kappa_{+} + 27 \,\nu^2 \,\big(2 + \,\kappa_{+}\big) \nl
&+ \,\nu \,\big(-114 + 23 \,\kappa_{+}\big)\big] \,v^4\big\} \,(nS)^2 + \,\Big(60 \,\nu \,\big\{\,\kappa_{-} -  \,\delta \,\kappa_{+} + 3 \,\nu \,\big[\,\kappa_{-} \nl
&+ \,\delta \,\big(2 + \,\kappa_{+}\big)\big]\big\} \,(nv)^3 - 6 \,\big\{\,\big(21 - 83 \,\nu + 93 \,\nu^2\big) \,\kappa_{-} + \,\delta \,\big[80 - 21 \,\kappa_{+} + 27 \,\nu^2 \,\big(2 + \,\kappa_{+}\big) \nl
&+ \,\nu \,\big(-174 + 41 \,\kappa_{+}\big)\big]\big\} \,(nv) \,v^2\Big) \,(vS) \,(n\Sigma) + \,\big\{-60 \,\nu \,\big[- \,\delta \,\kappa_{-} + \,\kappa_{+} + 6 \,\nu^2 \,\big(2 + \,\kappa_{+}\big)\nl
& - 2 \,\nu \,\big(7 + \,\kappa_{+}\big)\big] \,(nv)^3 + 6 \,\big[-80 - 21 \,\delta \,\kappa_{-} + 2 \,\nu \,\big(207 + 31 \,\delta \,\kappa_{-} - 52 \,\kappa_{+}\big) + 21 \,\kappa_{+}\nl
& + 54 \,\nu^3 \,\big(2 + \,\kappa_{+}\big) + \,\nu^2 \,\big(-596 - 33 \,\delta \,\kappa_{-} + 115 \,\kappa_{+}\big)\big] \,(nv) \,v^2\big\} \,(v\Sigma) \,(n\Sigma)\nl
& + \,\Big(210 \,\nu \,\big(-1 + 3 \,\nu\big) \,\big[\,\kappa_{-} -  \,\delta \,\big(2 + \,\kappa_{+}\big)\big] \,(nv)^4 + 60 \,\nu \,\big\{\,\kappa_{-} - 15 \,\nu \,\kappa_{-} \nl
&+ \,\delta \,\big[-10 -  \,\kappa_{+} + 15 \,\nu \,\big(2 + \,\kappa_{+}\big)\big]\big\} \,(nv)^2 \,v^2 \nl
&+ 6 \,\big\{\,\big(21 - 23 \,\nu - 27 \,\nu^2\big) \,\kappa_{-} + \,\delta \,\big[102 - 21 \,\kappa_{+} + 27 \,\nu^2 \,\big(2 + \,\kappa_{+}\big)\nl
& + \,\nu \,\big(-258 + 23 \,\kappa_{+}\big)\big]\big\} \,v^4\Big) \,(nS) \,(n\Sigma) + \,\big\{105 \,\nu \,\big(-1 + 3 \,\nu\big) \,\big[\,\delta \,\kappa_{-} -  \,\kappa_{+} + 2 \,\nu \,\big(2 + \,\kappa_{+}\big)\big] \,(nv)^4 \nl
&- 30 \,\nu \,\big(-1 + 15 \,\nu\big) \,\big[\,\delta \,\kappa_{-} -  \,\kappa_{+} + 2 \,\nu \,\big(2 + \,\kappa_{+}\big)\big] \,(nv)^2 \,v^2\nl
& - 3 \,\big[\,\nu \,\big(524 + 23 \,\delta \,\kappa_{-} - 65 \,\kappa_{+}\big) - 3 \,\big(32 + 7 \,\delta \,\kappa_{-} - 7 \,\kappa_{+}\big) + 54 \,\nu^3 \,\big(2 + \,\kappa_{+}\big) \nl
&+ \,\nu^2 \,\big(-804 + 27 \,\delta \,\kappa_{-} + 19 \,\kappa_{+}\big)\big] \,v^4\big\} \,(n\Sigma)^2 + \,\big\{30 \,\nu \,\big[-10 + 2 \,\delta \,\kappa_{-} + \,\kappa_{+} + 3 \,\nu \,\big(2 + \,\kappa_{+}\big)\big] \,(nv)^4\nl
& - 12 \,\big[\,\nu \,\big(-10 + 12 \,\delta \,\kappa_{-} - 3 \,\kappa_{+}\big) + 15 \,\nu^2 \,\big(2 + \,\kappa_{+}\big) + 3 \,\big(-4 -  \,\delta \,\kappa_{-} + \,\kappa_{+}\big)\big] \,(nv)^2 \,v^2\nl
& - 2 \,\big[78 + 6 \,\delta \,\kappa_{-} - 27 \,\kappa_{+} + 27 \,\nu^2 \,\big(2 + \,\kappa_{+}\big) + \,\nu \,\big(-154 - 18 \,\delta \,\kappa_{-} + 33 \,\kappa_{+}\big)\big] \,v^4\big\} \,\bS^2\nl
& + \,\Big(30 \,\nu \,\big[\,\kappa_{-} - 11 \,\nu \,\kappa_{-} + \,\big(-1 + 3 \,\nu\big) \,\delta \,\big(2 + \,\kappa_{+}\big)\big] \,(nv)^4 + \,\big\{36 \,\big(2 - 9 \,\nu + 21 \,\nu^2\big) \,\kappa_{-}\nl
& - 12 \,\delta \,\big[\,\nu \,\big(46 - 15 \,\kappa_{+}\big) + 6 \,\big(-4 + \,\kappa_{+}\big) + 15 \,\nu^2 \,\big(2 + \,\kappa_{+}\big)\big]\big\} \,(nv)^2 \,v^2\nl
& - 2 \,\big\{3 \,\big(11 - 25 \,\nu + 15 \,\nu^2\big) \,\kappa_{-} + \,\delta \,\big[150 - 33 \,\kappa_{+} + 27 \,\nu^2 \,\big(2 + \,\kappa_{+}\big) \nl
&+ \,\nu \,\big(-370 + 51 \,\kappa_{+}\big)\big]\big\} \,v^4\Big) \,(S\Sigma) + \,\big\{-15 \,\nu \,\big[- \,\delta \,\kappa_{-} + \,\nu \,\big(12 + 7 \,\delta \,\kappa_{-} - 9 \,\kappa_{+}\big)\nl
& + \,\kappa_{+} + 6 \,\nu^2 \,\big(2 + \,\kappa_{+}\big)\big] \,(nv)^4 + 6 \,\big[3 \,\nu^2 \,\big(68 + 13 \,\delta \,\kappa_{-} - 23 \,\kappa_{+}\big) + 6 \,\big(4 + \,\delta \,\kappa_{-} -  \,\kappa_{+}\big)\nl
& + 30 \,\nu^3 \,\big(2 + \,\kappa_{+}\big) + \,\nu \,\big(-128 - 21 \,\delta \,\kappa_{-} + 33 \,\kappa_{+}\big)\big] \,(nv)^2 \,v^2\nl
& + \,\big[-144 - 33 \,\delta \,\kappa_{-} + 3 \,\nu \,\big(260 + 21 \,\delta \,\kappa_{-} - 43 \,\kappa_{+}\big) + 33 \,\kappa_{+} + 54 \,\nu^3 \,\big(2 + \,\kappa_{+}\big)\nl
& + \,\nu^2 \,\big(-1172 - 9 \,\delta \,\kappa_{-} + 111 \,\kappa_{+}\big)\big] \,v^4\big\} \,\bSigma^2 \nl
& + \,(vS) \,\Big[\,\Big(-12 \,\big\{\,\big(-3 + 7 \,\nu + 21 \,\nu^2\big) \,\kappa_{-} + \,\delta \,\big[-8 -  \,\nu \,\big(-2 + \,\kappa_{+}\big) \nl
&+ 3 \,\kappa_{+} + 3 \,\nu^2 \,\big(2 + \,\kappa_{+}\big)\big]\big\} \,(nv)^2 + 4 \,\big\{6 \,\big(3 - 13 \,\nu + 18 \,\nu^2\big) \,\kappa_{-} \nl
&+ \,\delta \,\big[75 - 18 \,\kappa_{+} + 2 \,\nu \,\big(-89 + 21 \,\kappa_{+}\big)\big]\big\} \,v^2\Big) \,(v\Sigma) \Big]\,,\nl e_8^1&=4 \,\big[\,\nu \,\big(-4 + 3 \,\delta \,\kappa_{-} - 41 \,\kappa_{+}\big) + 5 \,\big(2 + \,\delta \,\kappa_{-} + \,\kappa_{+}\big)\big] \,(vS)^2 \nl
& + 8 \,\big\{-6 \,\big(-2 + \,\nu\big) \,\nu \,\kappa_{-}  + \,\delta \,\big[5 + \,\nu \,\big(6 - 22 \,\kappa_{+}\big)\big]\big\} \,(vS) \,(v\Sigma)  \nl
&+ 4 \,\nu \,\big[20 + 17 \,\delta \,\kappa_{-} - 17 \,\kappa_{+} + \,\nu \,\big(-38 - 3 \,\delta \,\kappa_{-} + 47 \,\kappa_{+}\big)\big] \,(v\Sigma)^2 \nl
& + \,\big[72 \,\nu^2 \,\big(2 + \,\kappa_{+}\big) - 8 \,\big(5 + 21 \,\delta \,\kappa_{-} + 21 \,\kappa_{+}\big) + 2 \,\nu \,\big(930 + 15 \,\delta \,\kappa_{-} + 649 \,\kappa_{+}\big)\big] \,(nv) \,(vS) \,(nS) \nl
& + \,\big\{-2 \,\nu \,\big(149 + 48 \,\nu\big) \,\kappa_{-} + \,\delta \,\big[-56 + 36 \,\nu^2 \,\big(2 + \,\kappa_{+}\big) + \,\nu \,\big(824 + 634 \,\kappa_{+}\big)\big]\big\} \,(nv) \,(v\Sigma) \,(nS) \nl
& + \,\big\{-6 \,\big[26 - 43 \,\delta \,\kappa_{-} - 43 \,\kappa_{+} + 40 \,\nu^2 \,\big(2 + \,\kappa_{+}\big) + \,\nu \,\big(740 + 27 \,\delta \,\kappa_{-} + 340 \,\kappa_{+}\big)\big] \,(nv)^2 \nl
& - 2 \,\big[-54 + 46 \,\delta \,\kappa_{-} + \,\nu \,\big(-682 + 9 \,\delta \,\kappa_{-} - 348 \,\kappa_{+}\big) + 46 \,\kappa_{+} + 12 \,\nu^2 \,\big(2 + \,\kappa_{+}\big)\big] \,v^2\big\} \,(nS)^2  \nl
&+ \,\big\{-2 \,\nu \,\big(149 + 48 \,\nu\big) \,\kappa_{-} + \,\delta \,\big[-32 + 36 \,\nu^2 \,\big(2 + \,\kappa_{+}\big) + \,\nu \,\big(876 + 634 \,\kappa_{+}\big)\big]\big\} \,(nv) \,(vS) \,(n\Sigma) \nl
& - 2 \,\nu \,\big[84 + 233 \,\delta \,\kappa_{-} - 233 \,\kappa_{+} + 36 \,\nu^2 \,\big(2 + \,\kappa_{+}\big) + \,\nu \,\big(730 + 33 \,\delta \,\kappa_{-} + 601 \,\kappa_{+}\big)\big] \,(nv) \,(v\Sigma) \,(n\Sigma) \nl
& + \,\Big(\,\big\{6 \,\nu \,\big(141 + 148 \,\nu\big) \,\kappa_{-} - 6 \,\nu \,\delta \,\big[752 + 313 \,\kappa_{+} + 40 \,\nu \,\big(2 + \,\kappa_{+}\big)\big]\big\} \,(nv)^2  \nl
&+ \,\big\{2 \,\nu \,\big(-173 + 48 \,\nu\big) \,\kappa_{-} + \,\delta \,\big[64 - 24 \,\nu^2 \,\big(2 + \,\kappa_{+}\big) + 6 \,\nu \,\big(246 + 119 \,\kappa_{+}\big)\big]\big\} \,v^2\Big) \,(nS) \,(n\Sigma)  \nl
&+ \,\big\{3 \,\nu \,\big[-60 + 227 \,\delta \,\kappa_{-} - 227 \,\kappa_{+} + 80 \,\nu^2 \,\big(2 + \,\kappa_{+}\big) + 2 \,\nu \,\big(768 + 47 \,\delta \,\kappa_{-} + 266 \,\kappa_{+}\big)\big] \,(nv)^2 \nl
& + \,\nu \,\big[156 - 265 \,\delta \,\kappa_{-} + 2 \,\nu \,\big(-824 + 15 \,\delta \,\kappa_{-} - 372 \,\kappa_{+}\big) + 265 \,\kappa_{+} + 24 \,\nu^2 \,\big(2 + \,\kappa_{+}\big)\big] \,v^2\big\} \,(n\Sigma)^2  \nl
&+ \,\big\{2 \,\big[28 \,\nu^2 \,\big(2 + \,\kappa_{+}\big) - 5 \,\big(-2 + 3 \,\delta \,\kappa_{-} + 3 \,\kappa_{+}\big) + \,\nu \,\big(658 + 22 \,\delta \,\kappa_{-} + 209 \,\kappa_{+}\big)\big] \,(nv)^2  \nl
&+ \,\big[2 \,\nu \,\big(-302 + \,\delta \,\kappa_{-} - 110 \,\kappa_{+}\big) + 8 \,\nu^2 \,\big(2 + \,\kappa_{+}\big) + 4 \,\big(-11 + 6 \,\delta \,\kappa_{-} + 6 \,\kappa_{+}\big)\big] \,v^2\big\} \,\bS^2  \nl
&+ \,\Big(\,\big\{-2 \,\nu \,\big(127 + 116 \,\nu\big) \,\kappa_{-} + 2 \,\delta \,\big[16 + 28 \,\nu^2 \,\big(2 + \,\kappa_{+}\big) + \,\nu \,\big(702 + 187 \,\kappa_{+}\big)\big]\big\} \,(nv)^2  \nl
&+ 2 \,\big\{\,\big(63 - 8 \,\nu\big) \,\nu \,\kappa_{-} + \,\delta \,\big[-24 + 4 \,\nu^2 \,\big(2 + \,\kappa_{+}\big) -  \,\nu \,\big(338 + 111 \,\kappa_{+}\big)\big]\big\} \,v^2\Big) \,(S\Sigma)  \nl
&+ \,\big\{- \,\nu \,\big[-124 + 157 \,\delta \,\kappa_{-} - 157 \,\kappa_{+} + 56 \,\nu^2 \,\big(2 + \,\kappa_{+}\big) + 2 \,\nu \,\big(778 + 36 \,\delta \,\kappa_{-} + 151 \,\kappa_{+}\big)\big] \,(nv)^2  \nl
&-  \,\nu \,\big[108 - 87 \,\delta \,\kappa_{-} + \,\nu \,\big(-808 + 6 \,\delta \,\kappa_{-} - 228 \,\kappa_{+}\big) + 87 \,\kappa_{+} + 8 \,\nu^2 \,\big(2 + \,\kappa_{+}\big)\big] \,v^2\big\} \,\bSigma^2\,,\nl
e_8^2&=\,\big[72 - 414 \,\delta \,\kappa_{-} - 120 \,\kappa_{+} + 94 \,\nu \,\big(50 + 17 \,\kappa_{+}\big)\big] \,(nS)^2 \nl
& + 2 \,\big\{\,\big(-147 + 29 \,\nu\big) \,\kappa_{-} + \,\delta \,\big[-70 + 147 \,\kappa_{+} + \,\nu \,\big(2322 + 799 \,\kappa_{+}\big)\big]\big\} \,(nS) \,(n\Sigma) \nl
&+ \,\big[147 \,\big(- \,\delta \,\kappa_{-} + \,\kappa_{+}\big) + \,\nu \,\big(420 - 385 \,\delta \,\kappa_{-} + 91 \,\kappa_{+}\big) - 2 \,\nu^2 \,\big(2280 + 799 \,\kappa_{+}\big)\big] \,(n\Sigma)^2  \nl
&+ 2 \,\big[72 + 69 \,\delta \,\kappa_{-} + 20 \,\kappa_{+} -  \,\nu \,\big(662 + 341 \,\kappa_{+}\big)\big] \,\bS^2 \nl
& + \,\big\{2 \,\big(49 + 65 \,\nu\big) \,\kappa_{-} - 2 \,\delta \,\big[-70 + 49 \,\kappa_{+} + \,\nu \,\big(634 + 341 \,\kappa_{+}\big)\big]\big\} \,(S\Sigma)  \nl
&+ \,\big[7 \,\nu \,\big(4 + 29 \,\delta \,\kappa_{-} - 15 \,\kappa_{+}\big) + 49 \,\big(\,\delta \,\kappa_{-} -  \,\kappa_{+}\big) + 2 \,\nu^2 \,\big(592 + 341 \,\kappa_{+}\big)\big] \,\bSigma^2\,.\nn
\end{align}
\endgroup
On the other hand, for the radiated flux in \eqref{fso}-\eqref{fss} we find
\begingroup
\allowdisplaybreaks
\scriptsize
\begin{align}
f_3^0&=-78 \,(nv)^2 \,(n, S, v) - 51 \,\delta \,(nv)^2 \,(n, \Sigma, v) + 80 \,(n, S, v) \,v^2 + 43 \,\delta \,(n, \Sigma, v) \,v^2\,,\nl
f_3^1&=2 \,(n, S, v) -  \,\delta \,(n, \Sigma, v)\,,\nl
f_5^0&=48 \,\big(187 - 262 \,\nu\big) \,(nv)^4 \,(n, S, v) + 3 \,\big(2647 - 3568 \,\nu\big) \,\delta \,(nv)^4 \,(n, \Sigma, v)\nl
& + 36 \,\big(-391 + 556 \,\nu\big) \,(nv)^2 \,(n, S, v) \,v^2 + 12 \,\big(-788 + 1207 \,\nu\big) \,\delta \,(nv)^2 \,(n, \Sigma, v) \,v^2 \nl
&+ \,\big(4828 - 7240 \,\nu\big) \,(n, S, v) \,v^4 + \,\big(2603 - 4160 \,\nu\big) \,\delta \,(n, \Sigma, v) \,v^4\,\nl
f_5^1&=22 \,\big(293 - 46 \,\nu\big) \,(nv)^2 \,(n, S, v) + \,\big(7327 - 2398 \,\nu\big) \,\delta \,(nv)^2 \,(n, \Sigma, v) \nl
&+ 2 \,\big(-3805 + 224 \,\nu\big) \,(n, S, v) \,v^2 + \,\big(-5387 + 994 \,\nu\big) \,\delta \,(n, \Sigma, v) \,v^2\,\nl
f_5^2&=-2 \,\big(472 + 195 \,\nu\big) \,(n, S, v) + \,\big(137 - 238 \,\nu\big) \,\delta \,(n, \Sigma, v)\,,\nl
f_7^0&=-10 \,\big(2939 - 26896 \,\nu + 6632 \,\nu^2\big) \,(nv)^6 \,(n, S, v) \nl
&+ 5 \,\big(12215 + 56391 \,\nu - 23312 \,\nu^2\big) \,\delta \,(nv)^6 \,(n, \Sigma, v) \nl
&+ 6 \,\big(11731 - 97552 \,\nu + 41208 \,\nu^2\big) \,(nv)^4 \,(n, S, v) \,v^2 \nl
&+ 3 \,\big(-28423 - 177451 \,\nu + 98604 \,\nu^2\big) \,\delta \,(nv)^4 \,(n, \Sigma, v) \,v^2 \nl
&- 6 \,\big(10485 - 67206 \,\nu + 44150 \,\nu^2\big) \,(nv)^2 \,(n, S, v) \,v^4 \nl
&+ 3 \,\big(4369 + 99655 \,\nu - 75750 \,\nu^2\big) \,\delta \,(nv)^2 \,(n, \Sigma, v) \,v^4 \nl
&+ 2 \,\big(11237 - 43924 \,\nu + 41766 \,\nu^2\big) \,(n, S, v) \,v^6 + \,\big(8847 - 49916 \,\nu + 50198 \,\nu^2\big) \,\delta \,(n, \Sigma, v) \,v^6\,,\nl
f_7^1&=-2 \,\big(36959 - 204812 \,\nu + 94378 \,\nu^2\big) \,(nv)^4 \,(n, S, v) \nl
&+ \,\big(-140184 + 619741 \,\nu - 243534 \,\nu^2\big) \,\delta \,(nv)^4 \,(n, \Sigma, v)\nl
& + 2 \,\big(87510 - 279467 \,\nu + 111074 \,\nu^2\big) \,(nv)^2 \,(n, S, v) \,v^2 \nl
&+ \,\big(177222 - 664927 \,\nu + 273594 \,\nu^2\big) \,\delta \,(nv)^2 \,(n, \Sigma, v) \,v^2\nl
& - 2 \,\big(44009 - 72691 \,\nu + 23602 \,\nu^2\big) \,(n, S, v) \,v^4 - 2 \,\big(33267 - 65918 \,\nu + 26015 \,\nu^2\big) \,\delta \,(n, \Sigma, v) \,v^4\,,\nl
f_7^2&=\,\big(165714 - 210143 \,\nu - 82280 \,\nu^2\big) \,(nv)^2 \,(n, S, v)\nl
& + \,\big(-169603 + 142753 \,\nu - 108405 \,\nu^2\big) \,\delta \,(nv)^2 \,(n, \Sigma, v) \nl
&+ \,\big(98244 + 77017 \,\nu + 23098 \,\nu^2\big) \,(n, S, v) \,v^2 + 2 \,\big(76462 - 19183 \,\nu + 14961 \,\nu^2\big) \,\delta \,(n, \Sigma, v) \,v^2\,,\nl
f_7^3&=3 \,\big(67927 + 32320 \,\nu - 2362 \,\nu^2\big) \,(n, S, v) + \,\big(21076 + 28209 \,\nu - 6021 \,\nu^2\big) \,\delta \,(n, \Sigma, v)\,\nl
f_4^0&=72 \,\big(2 + \,\kappa_{+}\big) \,(vS)^2 + 72 \,\big[- \,\kappa_{-} + \,\delta \,\big(2 + \,\kappa_{+}\big)\big] \,(vS) \,(v\Sigma) \nl
&+ \,\big[1 - 36 \,\delta \,\kappa_{-} + 36 \,\kappa_{+} - 72 \,\nu \,\big(2 + \,\kappa_{+}\big)\big] \,(v\Sigma)^2 - 348 \,\big(2 + \,\kappa_{+}\big) \,(nv) \,(vS) \,(nS)\nl
& + 174 \,\big[\,\kappa_{-} -  \,\delta \,\big(2 + \,\kappa_{+}\big)\big] \,(nv) \,(v\Sigma) \,(nS) + \,\big[816 \,\big(2 + \,\kappa_{+}\big) \,(nv)^2 - 504 \,\big(2 + \,\kappa_{+}\big) \,v^2\big] \,(nS)^2\nl
& + 174 \,\big[\,\kappa_{-} -  \,\delta \,\big(2 + \,\kappa_{+}\big)\big] \,(nv) \,(vS) \,(n\Sigma) \nl
&+ 6 \,\big[-1 + 29 \,\delta \,\kappa_{-} - 29 \,\kappa_{+} + 58 \,\nu \,\big(2 + \,\kappa_{+}\big)\big] \,(nv) \,(v\Sigma) \,(n\Sigma) \nl
&+ \,\big\{816 \,\big[- \,\kappa_{-} + \,\delta \,\big(2 + \,\kappa_{+}\big)\big] \,(nv)^2 + 504 \,\big[\,\kappa_{-} -  \,\delta \,\big(2 + \,\kappa_{+}\big)\big] \,v^2\big\} \,(nS) \,(n\Sigma) \nl
&+ \,\big\{\,\big[9 - 408 \,\delta \,\kappa_{-} + 408 \,\kappa_{+} - 816 \,\nu \,\big(2 + \,\kappa_{+}\big)\big] \,(nv)^2 + 252 \,\big[\,\delta \,\kappa_{-} -  \,\kappa_{+} + 2 \,\nu \,\big(2 + \,\kappa_{+}\big)\big] \,v^2\big\} \,(n\Sigma)^2 \nl
&+ \,\big[-156 \,\big(2 + \,\kappa_{+}\big) \,(nv)^2 + 144 \,\big(2 + \,\kappa_{+}\big) \,v^2\big] \,\bS^2 + \,\big\{156 \,\big[\,\kappa_{-} -  \,\delta \,\big(2 + \,\kappa_{+}\big)\big] \,(nv)^2\nl
&+ 144 \,\big[- \,\kappa_{-} + \,\delta \,\big(2 + \,\kappa_{+}\big)\big] \,v^2\big\} \,(S\Sigma) + \,\big\{\,\big[9 + 78 \,\delta \,\kappa_{-} - 78 \,\kappa_{+} + 156 \,\nu \,\big(2 + \,\kappa_{+}\big)\big] \,(nv)^2 \nl
&+ \,\big[3 - 72 \,\delta \,\kappa_{-} + 72 \,\kappa_{+} - 144 \,\nu \,\big(2 + \,\kappa_{+}\big)\big] \,v^2\big\} \,\bSigma^2\,,\nl
f_6^0&=\,\big\{\,\big[3897 + 3189 \,\delta \,\kappa_{-} - 4557 \,\kappa_{+} + 2550 \,\nu \,\big(2 + \,\kappa_{+}\big)\big] \,(nv)^2\nl
& + \,\big[-3239 - 834 \,\delta \,\kappa_{-} - 162 \,\kappa_{+} + 60 \,\nu \,\big(2 + \,\kappa_{+}\big)\big] \,v^2\big\} \,(vS)^2 \nl
&+ \,\Big(6 \,\big\{\,\big(1291 - 2551 \,\nu\big) \,\kappa_{-} + \,\delta \,\big[1921 - 1291 \,\kappa_{+} + 425 \,\nu \,\big(2 + \,\kappa_{+}\big)\big]\big\} \,(nv)^2 \nl
&+ \,\big\{84 \,\big(-8 + 39 \,\nu\big) \,\kappa_{-} + 2 \,\delta \,\big[-2047 + 336 \,\kappa_{+} + 30 \,\nu \,\big(2 + \,\kappa_{+}\big)\big]\big\} \,v^2\Big) \,(vS) \,(v\Sigma) \nl
&+ \,\big\{\,\big[7062 + 3873 \,\delta \,\kappa_{-} - 6 \,\nu \,\big(3036 + 744 \,\delta \,\kappa_{-} - 2035 \,\kappa_{+}\big) - 3873 \,\kappa_{+} - 2550 \,\nu^2 \,\big(2 + \,\kappa_{+}\big)\big] \,(nv)^2\nl
& + \,\big[-1121 - 336 \,\delta \,\kappa_{-} + 2 \,\nu \,\big(2447 + 402 \,\delta \,\kappa_{-} - 738 \,\kappa_{+}\big) + 336 \,\kappa_{+} - 60 \,\nu^2 \,\big(2 + \,\kappa_{+}\big)\big] \,v^2\big\} \,(v\Sigma)^2 \nl
&+ \,\big\{-12 \,\big[-686 + 823 \,\delta \,\kappa_{-} - 2159 \,\kappa_{+} + 1216 \,\nu \,\big(2 + \,\kappa_{+}\big)\big] \,(nv)^3 \nl
&+ \,\big[-90 + 4761 \,\delta \,\kappa_{-} - 9945 \,\kappa_{+} + 2046 \,\nu \,\big(2 + \,\kappa_{+}\big)\big] \,(nv) \,v^2\big\} \,(vS) \,(nS) \nl
&+ \,\Big(-12 \,\big\{7 \,\big(213 - 322 \,\nu\big) \,\kappa_{-} + \,\delta \,\big[10 - 1491 \,\kappa_{+} + 608 \,\nu \,\big(2 + \,\kappa_{+}\big)\big]\big\} \,(nv)^3\nl
& + \,\big\{57 \,\big(129 - 185 \,\nu\big) \,\kappa_{-} + 3 \,\delta \,\big[-854 - 2451 \,\kappa_{+} + 341 \,\nu \,\big(2 + \,\kappa_{+}\big)\big]\big\} \,(nv) \,v^2\Big) \,(v\Sigma) \,(nS)\nl
& + \,\big\{15 \,\big[-2719 + 431 \,\delta \,\kappa_{-} - 3019 \,\kappa_{+} + 2384 \,\nu \,\big(2 + \,\kappa_{+}\big)\big] \,(nv)^4 \nl
&- 6 \,\big[-5435 + 677 \,\delta \,\kappa_{-} - 6803 \,\kappa_{+} + 5248 \,\nu \,\big(2 + \,\kappa_{+}\big)\big] \,(nv)^2 \,v^2\nl
& + \,\big[-3273 + 285 \,\delta \,\kappa_{-} - 7389 \,\kappa_{+} + 6816 \,\nu \,\big(2 + \,\kappa_{+}\big)\big] \,v^4\big\} \,(nS)^2 \nl
&+ \,\Big(-3 \,\big\{28 \,\big(213 - 322 \,\nu\big) \,\kappa_{-} + \,\delta \,\big[2567 - 5964 \,\kappa_{+} + 2432 \,\nu \,\big(2 + \,\kappa_{+}\big)\big]\big\} \,(nv)^3\nl
&+ \,\big\{57 \,\big(129 - 185 \,\nu\big) \,\kappa_{-} + 3 \,\delta \,\big[1365 - 2451 \,\kappa_{+} + 341 \,\nu \,\big(2 + \,\kappa_{+}\big)\big]\big\} \,(nv) \,v^2\Big) \,(vS) \,(n\Sigma) \nl
&+ \,\big\{3 \,\big[-5589 - 5964 \,\delta \,\kappa_{-} + \,\nu \,\big(8143 + 5724 \,\delta \,\kappa_{-} - 17652 \,\kappa_{+}\big)\nl
& + 5964 \,\kappa_{+} + 4864 \,\nu^2 \,\big(2 + \,\kappa_{+}\big)\big] \,(nv)^3\nl
& - 3 \,\big[-1297 - 2451 \,\delta \,\kappa_{-} + \,\nu \,\big(1299 + 1928 \,\delta \,\kappa_{-} - 6830 \,\kappa_{+}\big) \nl
&+ 2451 \,\kappa_{+} + 682 \,\nu^2 \,\big(2 + \,\kappa_{+}\big)\big] \,(nv) \,v^2\big\} \,(v\Sigma) \,(n\Sigma) \nl
&+ \,\Big(30 \,\big[\,\big(1725 - 2054 \,\nu\big) \,\kappa_{-} + 1192 \,\nu \,\delta \,\big(2 + \,\kappa_{+}\big) - 3 \,\delta \,\big(394 + 575 \,\kappa_{+}\big)\big] \,(nv)^4\nl
& - 24 \,\big[17 \,\big(110 - 117 \,\nu\big) \,\kappa_{-} + 1312 \,\nu \,\delta \,\big(2 + \,\kappa_{+}\big) - 374 \,\delta \,\big(4 + 5 \,\kappa_{+}\big)\big] \,(nv)^2 \,v^2 \nl
&+ 6 \,\big\{\,\big(1279 - 1326 \,\nu\big) \,\kappa_{-} + \,\delta \,\big[-722 - 1279 \,\kappa_{+} + 1136 \,\nu \,\big(2 + \,\kappa_{+}\big)\big]\big\} \,v^4\Big) \,(nS) \,(n\Sigma)\nl
& + \,\big\{-15 \,\big[-614 - 1725 \,\delta \,\kappa_{-} + \,\nu \,\big(-2032 + 1623 \,\delta \,\kappa_{-}\nl
& - 5073 \,\kappa_{+}\big) + 1725 \,\kappa_{+} + 2384 \,\nu^2 \,\big(2 + \,\kappa_{+}\big)\big] \,(nv)^4 \nl
&+ 6 \,\big[\,\nu \,\big(-6634 + 3301 \,\delta \,\kappa_{-} - 10781 \,\kappa_{+}\big) - 4 \,\big(169 + 935 \,\delta \,\kappa_{-} \nl
&- 935 \,\kappa_{+}\big) + 5248 \,\nu^2 \,\big(2 + \,\kappa_{+}\big)\big] \,(nv)^2 \,v^2\nl
& + 3 \,\big[515 + 1279 \,\delta \,\kappa_{-} - 1279 \,\kappa_{+} - 2272 \,\nu^2 \,\big(2 + \,\kappa_{+}\big) \nl
&+ \,\nu \,\big(1814 - 1231 \,\delta \,\kappa_{-} + 3789 \,\kappa_{+}\big)\big] \,v^4\big\} \,(n\Sigma)^2\nl
& + \,\big\{3 \,\big[5906 + 379 \,\delta \,\kappa_{-} + 2153 \,\kappa_{+} - 2352 \,\nu \,\big(2 + \,\kappa_{+}\big)\big] \,(nv)^4 \nl
&+ 12 \,\big[-2166 - 108 \,\delta \,\kappa_{-} - 731 \,\kappa_{+} + 747 \,\nu \,\big(2 + \,\kappa_{+}\big)\big] \,(nv)^2 \,v^2 \nl
&+ 3 \,\big[3046 + 61 \,\delta \,\kappa_{-} + 839 \,\kappa_{+} - 764 \,\nu \,\big(2 + \,\kappa_{+}\big)\big] \,v^4\big\} \,\bS^2\nl
& + \,\Big(-6 \,\big\{\,\big(887 - 418 \,\nu\big) \,\kappa_{-} + \,\delta \,\big[-4221 - 887 \,\kappa_{+} + 1176 \,\nu \,\big(2 + \,\kappa_{+}\big)\big]\big\} \,(nv)^4 \nl
&+ 12 \,\big\{7 \,\big(89 - 45 \,\nu\big) \,\kappa_{-} + \,\delta \,\big[-2903 - 623 \,\kappa_{+} + 747 \,\nu \,\big(2 + \,\kappa_{+}\big)\big]\big\} \,(nv)^2 \,v^2 \nl
&+ \,\big\{6 \,\big(-389 + 260 \,\nu\big) \,\kappa_{-} + \,\delta \,\big[10406 + 2334 \,\kappa_{+} - 2292 \,\nu \,\big(2 + \,\kappa_{+}\big)\big]\big\} \,v^4\Big) \,(S\Sigma)\nl
& + \,\big\{3 \,\big[1619 - 887 \,\delta \,\kappa_{-} + \,\nu \,\big(-10798 + 797 \,\delta \,\kappa_{-}\nl
& - 2571 \,\kappa_{+}\big) + 887 \,\kappa_{+} + 2352 \,\nu^2 \,\big(2 + \,\kappa_{+}\big)\big] \,(nv)^4\nl
& - 3 \,\big[2381 - 1246 \,\delta \,\kappa_{-} + 2 \,\nu \,\big(-7193 + 531 \,\delta \,\kappa_{-} - 1777 \,\kappa_{+}\big) + 1246 \,\kappa_{+} + 2988 \,\nu^2 \,\big(2 + \,\kappa_{+}\big)\big] \,(nv)^2 \,v^2\nl
& + \,\big[1898 - 1167 \,\delta \,\kappa_{-} + \,\nu \,\big(-11668 + 963 \,\delta \,\kappa_{-} - 3297 \,\kappa_{+}\big) \nl
&+ 1167 \,\kappa_{+} + 2292 \,\nu^2 \,\big(2 + \,\kappa_{+}\big)\big] \,v^4\big\} \,\bSigma^2\,,\nl
f_6^1&=-2 \,\big[6810 - 1166 \,\delta \,\kappa_{-} + 3663 \,\kappa_{+} + 42 \,\nu \,\big(2 + \,\kappa_{+}\big)\big] \,(vS)^2 \nl
&- 2 \,\big\{\,\big(-4829 + 4622 \,\nu\big) \,\kappa_{-} + \,\delta \,\big[4620 + 4829 \,\kappa_{+} + 42 \,\nu \,\big(2 + \,\kappa_{+}\big)\big]\big\} \,(vS) \,(v\Sigma) \nl
&+ \,\big[2392 + 4829 \,\delta \,\kappa_{-} - 4829 \,\kappa_{+} + 84 \,\nu^2 \,\big(2 + \,\kappa_{+}\big) + \,\nu \,\big(4904 - 2290 \,\delta \,\kappa_{-} + 11948 \,\kappa_{+}\big)\big] \,(v\Sigma)^2\nl
& + 6 \,\big[10250 - 1706 \,\delta \,\kappa_{-} + 5338 \,\kappa_{+} + 161 \,\nu \,\big(2 + \,\kappa_{+}\big)\big] \,(nv) \,(vS) \,(nS) \nl
&+ \,\big[9 \,\big(-2348 + 2221 \,\nu\big) \,\kappa_{-} + 483 \,\nu \,\delta \,\big(2 + \,\kappa_{+}\big) + 4 \,\delta \,\big(6343 + 5283 \,\kappa_{+}\big)\big] \,(nv) \,(v\Sigma) \,(nS)\nl
& + \,\big\{6 \,\big[-25044 + 2399 \,\delta \,\kappa_{-} - 11299 \,\kappa_{+} + 169 \,\nu \,\big(2 + \,\kappa_{+}\big)\big] \,(nv)^2 \nl
&+ \,\big[91376 - 6190 \,\delta \,\kappa_{-} + 38868 \,\kappa_{+} - 996 \,\nu \,\big(2 + \,\kappa_{+}\big)\big] \,v^2\big\} \,(nS)^2\nl
& + \,\big[9 \,\big(-2348 + 2221 \,\nu\big) \,\kappa_{-} + 483 \,\nu \,\delta \,\big(2 + \,\kappa_{+}\big) + 4 \,\delta \,\big(5804 + 5283 \,\kappa_{+}\big)\big] \,(nv) \,(vS) \,(n\Sigma) \nl
&+ \,\big[\,\nu \,\big(-34902 + 9753 \,\delta \,\kappa_{-} - 52017 \,\kappa_{+}\big) - 4 \,\big(1226 + 5283 \,\delta \,\kappa_{-}\nl
& - 5283 \,\kappa_{+}\big) - 966 \,\nu^2 \,\big(2 + \,\kappa_{+}\big)\big] \,(nv) \,(v\Sigma) \,(n\Sigma) \nl
&+ \,\Big(6 \,\big\{9 \,\big(1522 - 1085 \,\nu\big) \,\kappa_{-} + \,\delta \,\big[-23537 - 13698 \,\kappa_{+} + 169 \,\nu \,\big(2 + \,\kappa_{+}\big)\big]\big\} \,(nv)^2 \nl
&+ \,\big\{2 \,\big(-22529 + 12878 \,\nu\big) \,\kappa_{-} + \,\delta \,\big[91858 + 45058 \,\kappa_{+} - 996 \,\nu \,\big(2 + \,\kappa_{+}\big)\big]\big\} \,v^2\Big) \,(nS) \,(n\Sigma)\nl
& + \,\big\{-3 \,\big[934 - 13698 \,\delta \,\kappa_{-} + \,\nu \,\big(-44112 + 4967 \,\delta \,\kappa_{-} \nl
&- 32363 \,\kappa_{+}\big) + 13698 \,\kappa_{+} + 338 \,\nu^2 \,\big(2 + \,\kappa_{+}\big)\big] \,(nv)^2 \nl
&+ \,\big[4850 - 22529 \,\delta \,\kappa_{-} + \,\nu \,\big(-92766 + 6688 \,\delta \,\kappa_{-} - 51746 \,\kappa_{+}\big)\nl
& + 22529 \,\kappa_{+} + 996 \,\nu^2 \,\big(2 + \,\kappa_{+}\big)\big] \,v^2\big\} \,(n\Sigma)^2 \nl
&+ \,\big\{\,\big[24436 - 1386 \,\delta \,\kappa_{-} + 11922 \,\kappa_{+} - 660 \,\nu \,\big(2 + \,\kappa_{+}\big)\big] \,(nv)^2 \nl
&+ 2 \,\big[643 \,\delta \,\kappa_{-} + 180 \,\nu \,\big(2 + \,\kappa_{+}\big) - 7 \,\big(1494 + 751 \,\kappa_{+}\big)\big] \,v^2\big\} \,\bS^2\nl
& + \,\Big(-12 \,\big\{\,\big(1109 - 517 \,\nu\big) \,\kappa_{-} + \,\delta \,\big[-2299 - 1109 \,\kappa_{+} + 55 \,\nu \,\big(2 + \,\kappa_{+}\big)\big]\big\} \,(nv)^2 \nl
&+ \,\big\{8 \,\big(1475 - 688 \,\nu\big) \,\kappa_{-} + 4 \,\delta \,\big[-6109 - 2950 \,\kappa_{+} + 90 \,\nu \,\big(2 + \,\kappa_{+}\big)\big]\big\} \,v^2\Big) \,(S\Sigma) \nl
&+ \,\big\{\,\big[604 - 6654 \,\delta \,\kappa_{-} + 4 \,\nu \,\big(-7717 + 429 \,\delta \,\kappa_{-} - 3756 \,\kappa_{+}\big) + 6654 \,\kappa_{+} + 660 \,\nu^2 \,\big(2 + \,\kappa_{+}\big)\big] \,(nv)^2 \nl
&- 2 \,\big[1102 - 2950 \,\delta \,\kappa_{-} + \,\nu \,\big(-14072 + 733 \,\delta \,\kappa_{-} - 6633 \,\kappa_{+}\big) \nl
&+ 2950 \,\kappa_{+} + 180 \,\nu^2 \,\big(2 + \,\kappa_{+}\big)\big] \,v^2\big\} \,\bSigma^2\,\nl
f_6^2&=3 \,\big[-39 - 2 \,\delta \,\kappa_{-} - 18 \,\kappa_{+} + 72 \,\nu \,\big(2 + \,\kappa_{+}\big)\big] \,(nS)^2 + \,\Big\{48 \,\big(1 - 4 \,\nu\big) \,\kappa_{-} \nl
&+ \,\delta \,\big[-98 - 48 \,\kappa_{+} + 216 \,\nu \,\big(2 + \,\kappa_{+}\big)\big]\Big\} \,(nS) \,(n\Sigma) - 2 \,\big[1 - 12 \,\delta \,\kappa_{-} + \,\nu \,\big(-58 + 51 \,\delta \,\kappa_{-} - 75 \,\kappa_{+}\big) \nl
&+ 12 \,\kappa_{+} + 108 \,\nu^2 \,\big(2 + \,\kappa_{+}\big)\big] \,(n\Sigma)^2 + 2 \,\big[23 + \,\delta \,\kappa_{-} + 9 \,\kappa_{+} - 36 \,\nu \,\big(2 + \,\kappa_{+}\big)\big] \,\bS^2 \nl
&- 4 \,\Big\{4 \,\big(1 - 4 \,\nu\big) \,\kappa_{-} + \,\delta \,\big[-7 - 4 \,\kappa_{+} + 18 \,\nu \,\big(2 + \,\kappa_{+}\big)\big]\Big\} \,(S\Sigma) \nl
&+ \,\big[3 - 8 \,\delta \,\kappa_{-} + \,\nu \,\big(-48 + 34 \,\delta \,\kappa_{-} - 50 \,\kappa_{+}\big) + 8 \,\kappa_{+} + 72 \,\nu^2 \,\big(2 + \,\kappa_{+}\big)\big] \,\bSigma^2\,\nl
f_8^0&=\,\big\{-3 \,\big[-88255 + 204790 \,\delta \,\kappa_{-} + 21058 \,\kappa_{+} + 14048 \,\nu^2 \,\big(2 + \,\kappa_{+}\big)\nl
& + \,\nu \,\big(47364 - 135196 \,\delta \,\kappa_{-} + 25148 \,\kappa_{+}\big)\big] \,(nv)^4 + 3 \,\big[-52309 + 163221 \,\delta \,\kappa_{-} + 39339 \,\kappa_{+} \nl
&+ 24630 \,\nu^2 \,\big(2 + \,\kappa_{+}\big) - 2 \,\nu \,\big(-8060 + 59661 \,\delta \,\kappa_{-} + 19254 \,\kappa_{+}\big)\big] \,(nv)^2 \,v^2 \nl
&- 6 \,\big[12058 + 7347 \,\delta \,\kappa_{-} + 3273 \,\kappa_{+} + 4872 \,\nu^2 \,\big(2 + \,\kappa_{+}\big)\nl
& - 6 \,\nu \,\big(1883 + 1279 \,\delta \,\kappa_{-} + 1941 \,\kappa_{+}\big)\big] \,v^4\big\} \,(vS)^2 \nl
&+ \,\Big(-6 \,\big\{2 \,\big(45933 - 244876 \,\nu + 131684 \,\nu^2\big) \,\kappa_{-} + \,\delta \,\big[90509 - 91866 \,\kappa_{+} + 7024 \,\nu^2 \,\big(2 + \,\kappa_{+}\big) \nl
&+ \,\nu \,\big(-153137 + 80172 \,\kappa_{+}\big)\big]\big\} \,(nv)^4 \nl
&+ 6 \,\big\{3 \,\big(20647 - 122283 \,\nu + 75443 \,\nu^2\big) \,\kappa_{-} + \,\delta \,\big[70079 - 61941 \,\kappa_{+} + 12315 \,\nu^2 \,\big(2 + \,\kappa_{+}\big) \nl
&+ 3 \,\nu \,\big(-48145 + 13469 \,\kappa_{+}\big)\big]\big\} \,(nv)^2 \,v^2 - 12 \,\big\{3 \,\big(679 - 4236 \,\nu + 4304 \,\nu^2\big) \,\kappa_{-} \nl
&+ \,\delta \,\big[10347 - 2037 \,\kappa_{+} + 2436 \,\nu^2 \,\big(2 + \,\kappa_{+}\big) - 2 \,\nu \,\big(5947 + 993 \,\kappa_{+}\big)\big]\big\} \,v^4\Big) \,(vS) \,(v\Sigma) \nl
&+ \,\big\{3 \,\big[-175013 - 91866 \,\delta \,\kappa_{-} + \,\nu \,\big(581226 + 284962 \,\delta \,\kappa_{-} - 468694 \,\kappa_{+}\big)\nl
& + 91866 \,\kappa_{+} + 14048 \,\nu^3 \,\big(2 + \,\kappa_{+}\big) + \,\nu^2 \,\big(-638641 - 128172 \,\delta \,\kappa_{-} + 288516 \,\kappa_{+}\big)\big] \,(nv)^4 \nl
&+ \,\big[9 \,\nu^2 \,\big(189985 + 35669 \,\delta \,\kappa_{-} - 62607 \,\kappa_{+}\big) - 18 \,\nu \,\big(72279 + 33938 \,\delta \,\kappa_{-} - 54585 \,\kappa_{+}\big) \nl
&+ 33 \,\big(10717 + 5631 \,\delta \,\kappa_{-} - 5631 \,\kappa_{+}\big) - 73890 \,\nu^3 \,\big(2 + \,\kappa_{+}\big)\big] \,(nv)^2 \,v^2\nl
& + 6 \,\big[\,\nu \,\big(31842 + 5361 \,\delta \,\kappa_{-} - 9435 \,\kappa_{+}\big) - 3 \,\big(1934 + 679 \,\delta \,\kappa_{-} - 679 \,\kappa_{+}\big) \nl
&- 2 \,\nu^2 \,\big(17741 + 2619 \,\delta \,\kappa_{-} - 633 \,\kappa_{+}\big) + 4872 \,\nu^3 \,\big(2 + \,\kappa_{+}\big)\big] \,v^4\big\} \,(v\Sigma)^2 \nl
&+ \,\Big(90 \,\big[-20466 + 22857 \,\delta \,\kappa_{-} + 3619 \,\kappa_{+} + 1232 \,\nu^2 \,\big(2 + \,\kappa_{+}\big) + \,\nu \,\big(35394 - 14094 \,\delta \,\kappa_{-} \nl
&+ 8402 \,\kappa_{+}\big)\big] \,(nv)^5 - 12 \,\big[-176174 + 171751 \,\delta \,\kappa_{-} + 50677 \,\kappa_{+} + 32588 \,\nu^2 \,\big(2 + \,\kappa_{+}\big) \nl
&+ \,\nu \,\big(333866 - 116344 \,\delta \,\kappa_{-} + 22292 \,\kappa_{+}\big)\big] \,(nv)^3 \,v^2 + 3 \,\big\{-132202 + 118101 \,\delta \,\kappa_{-}\nl
& + 46851 \,\kappa_{+} + 89034 \,\nu^2 \,\big(2 + \,\kappa_{+}\big) - 2 \,\nu \,\big[54081 \,\delta \,\kappa_{-} + 26 \,\big(-7777 + 492 \,\kappa_{+}\big)\big]\big\} \,(nv) \,v^4\Big) \,(vS) \,(nS)\nl
& + \,\Big(90 \,\big\{\,\big(9619 - 56962 \,\nu + 27572 \,\nu^2\big) \,\kappa_{-} + \,\delta \,\big[1766 - 9619 \,\kappa_{+} + 616 \,\nu^2 \,\big(2 + \,\kappa_{+}\big) \nl
&+ 2 \,\nu \,\big(6153 + 5624 \,\kappa_{+}\big)\big]\big\} \,(nv)^5 - 12 \,\big\{\,\big(60537 - 412820 \,\nu + 216394 \,\nu^2\big) \,\kappa_{-} + \,\delta \,\big[210 - 60537 \,\kappa_{+} \nl
&+ 16294 \,\nu^2 \,\big(2 + \,\kappa_{+}\big) + 2 \,\nu \,\big(79897 + 34659 \,\kappa_{+}\big)\big]\big\} \,(nv)^3 \,v^2 + \,\big\{9 \,\big(11875 - 92497 \,\nu + 57269 \,\nu^2\big) \,\kappa_{-}\nl
& + 3 \,\delta \,\big[19862 - 35625 \,\kappa_{+} + 44517 \,\nu^2 \,\big(2 + \,\kappa_{+}\big) + 3 \,\nu \,\big(88346 + 13763 \,\kappa_{+}\big)\big]\big\} \,(nv) \,v^4\Big) \,(v\Sigma) \,(nS) \nl
&+ \,\big\{30 \,\big[92054 - 51639 \,\delta \,\kappa_{-} + 12 \,\nu \,\big(-21737 + 2383 \,\delta \,\kappa_{-} - 7117 \,\kappa_{+}\big) + 3159 \,\kappa_{+} \nl
&+ 17664 \,\nu^2 \,\big(2 + \,\kappa_{+}\big)\big] \,(nv)^6 - 30 \,\big[145074 - 60071 \,\delta \,\kappa_{-} + 2 \,\nu \,\big(-217150 + 17845 \,\delta \,\kappa_{-}\nl
& - 66746 \,\kappa_{+}\big) + 5839 \,\kappa_{+} + 38456 \,\nu^2 \,\big(2 + \,\kappa_{+}\big)\big] \,(nv)^4 \,v^2 + 6 \,\big[316786 - 73665 \,\delta \,\kappa_{-} \nl
&+ 4 \,\nu \,\big(-255289 + 13716 \,\delta \,\kappa_{-} - 87213 \,\kappa_{+}\big) + 50505 \,\kappa_{+} + 144912 \,\nu^2 \,\big(2 + \,\kappa_{+}\big)\big] \,(nv)^2 \,v^4 \nl
&- 6 \,\big[31638 - 1905 \,\delta \,\kappa_{-} + \,\nu \,\big(-81524 + 2538 \,\delta \,\kappa_{-} - 58068 \,\kappa_{+}\big) \nl
&+ 19521 \,\kappa_{+} + 39720 \,\nu^2 \,\big(2 + \,\kappa_{+}\big)\big] \,v^6\big\} \,(nS)^2 \nl
&+ \,\Big(90 \,\big\{\,\big(9619 - 56962 \,\nu + 27572 \,\nu^2\big) \,\kappa_{-} + \,\delta \,\big[7744 - 9619 \,\kappa_{+} + 616 \,\nu^2 \,\big(2 + \,\kappa_{+}\big) \nl
&+ 4 \,\nu \,\big(-295 + 2812 \,\kappa_{+}\big)\big]\big\} \,(nv)^5 - 3 \,\big\{4 \,\big(60537 - 412820 \,\nu + 216394 \,\nu^2\big) \,\kappa_{-}\nl
& + \,\delta \,\big[182127 - 242148 \,\kappa_{+} + 65176 \,\nu^2 \,\big(2 + \,\kappa_{+}\big) + \,\nu \,\big(58154 + 277272 \,\kappa_{+}\big)\big]\big\} \,(nv)^3 \,v^2\nl
& + \,\big\{9 \,\big(11875 - 92497 \,\nu + 57269 \,\nu^2\big) \,\kappa_{-} + 3 \,\delta \,\big[35069 - 35625 \,\kappa_{+} + 44517 \,\nu^2 \,\big(2 + \,\kappa_{+}\big) \nl
&+ \,\nu \,\big(77244 + 41289 \,\kappa_{+}\big)\big]\big\} \,(nv) \,v^4\Big) \,(vS) \,(n\Sigma) + \,\big\{-90 \,\big[-14140 - 9619 \,\delta \,\kappa_{-}\nl
& + \,\nu \,\big(45414 + 34105 \,\delta \,\kappa_{-} - 53343 \,\kappa_{+}\big) - 2 \,\nu^2 \,\big(7778 + 6739 \,\delta \,\kappa_{-} - 17987 \,\kappa_{+}\big) + 9619 \,\kappa_{+} \nl
&+ 1232 \,\nu^3 \,\big(2 + \,\kappa_{+}\big)\big] \,(nv)^5 + 3 \,\big[-321287 - 242148 \,\delta \,\kappa_{-} + \,\nu \,\big(1052063 + 964276 \,\delta \,\kappa_{-} - 1448572 \,\kappa_{+}\big)\nl
& + 242148 \,\kappa_{+} + 130352 \,\nu^3 \,\big(2 + \,\kappa_{+}\big) + \,\nu^2 \,\big(-17459 - 400200 \,\delta \,\kappa_{-} + 954744 \,\kappa_{+}\big)\big] \,(nv)^3 \,v^2 \nl
&- 3 \,\big[-45701 - 35625 \,\delta \,\kappa_{-} + 15 \,\nu \,\big(9357 + 10626 \,\delta \,\kappa_{-} - 15376 \,\kappa_{+}\big) + 35625 \,\kappa_{+} + 89034 \,\nu^3 \,\big(2 + \,\kappa_{+}\big) \nl
&+ \,\nu^2 \,\big(260391 - 63645 \,\delta \,\kappa_{-} + 146223 \,\kappa_{+}\big)\big] \,(nv) \,v^4\big\} \,(v\Sigma) \,(n\Sigma) \nl
&+ \,\Big(60 \,\big\{-3 \,\big(9133 - 53426 \,\nu + 22008 \,\nu^2\big) \,\kappa_{-}\nl
& + \,\delta \,\big[5136 + 27399 \,\kappa_{+} + 8832 \,\nu^2 \,\big(2 + \,\kappa_{+}\big) - 4 \,\nu \,\big(29311 + 14250 \,\kappa_{+}\big)\big]\big\} \,(nv)^6\nl
& - 60 \,\big\{\,\big(-32955 + 204733 \,\nu - 90608 \,\nu^2\big) \,\kappa_{-} + \,\delta \,\big[19228 \,\nu^2 \,\big(2 + \,\kappa_{+}\big) + 5 \,\big(4094 + 6591 \,\kappa_{+}\big)\nl
& - 3 \,\nu \,\big(70417 + 28197 \,\kappa_{+}\big)\big]\big\} \,(nv)^4 \,v^2 \nl
&+ 12 \,\big\{-3 \,\big(20695 - 116396 \,\nu + 60728 \,\nu^2\big) \,\kappa_{-} + \,\delta \,\big[72796 + 62085 \,\kappa_{+} \nl
&+ 72456 \,\nu^2 \,\big(2 + \,\kappa_{+}\big) - 6 \,\nu \,\big(88651 + 33643 \,\kappa_{+}\big)\big]\big\} \,(nv)^2 \,v^4 \nl
&+ \,\big\{36 \,\big(3571 - 11371 \,\nu + 8312 \,\nu^2\big) \,\kappa_{-}- 12 \,\delta \,\big[11822 + 10713 \,\kappa_{+} + 19860 \,\nu^2 \,\big(2 + \,\kappa_{+}\big) \nl
&- 13 \,\nu \,\big(3467 + 2331 \,\kappa_{+}\big)\big]\big\} \,v^6\Big) \,(nS) \,(n\Sigma)\nl
& + \,\big\{-30 \,\big[20690 + 27399 \,\delta \,\kappa_{-} + 2 \,\nu^2 \,\big(-104335 + 18714 \,\delta \,\kappa_{-} - 75714 \,\kappa_{+}\big) \nl
&- 27399 \,\kappa_{+} + 17664 \,\nu^3 \,\big(2 + \,\kappa_{+}\big) + \,\nu \,\big(-46172 - 108639 \,\delta \,\kappa_{-} + 163437 \,\kappa_{+}\big)\big] \,(nv)^6 \nl
&+ 30 \,\big[12682 + 32955 \,\delta \,\kappa_{-} + 6 \,\nu^2 \,\big(-69356 + 9153 \,\delta \,\kappa_{-} - 37350 \,\kappa_{+}\big) - 32955 \,\kappa_{+}\nl
& + 38456 \,\nu^3 \,\big(2 + \,\kappa_{+}\big) + \,\nu \,\big(3736 - 144662 \,\delta \,\kappa_{-} + 210572 \,\kappa_{+}\big)\big] \,(nv)^4 \,v^2\nl
& - 6 \,\big[1658 + 62085 \,\delta \,\kappa_{-} + 6 \,\nu^2 \,\big(-187653 + 21220 \,\delta \,\kappa_{-} - 88506 \,\kappa_{+}\big) - 62085 \,\kappa_{+}\nl
& + 144912 \,\nu^3 \,\big(2 + \,\kappa_{+}\big) + \,\nu \,\big(159576 - 275523 \,\delta \,\kappa_{-} + 399693 \,\kappa_{+}\big)\big] \,(nv)^2 \,v^4\nl
& + 6 \,\big[2876 + 10713 \,\delta \,\kappa_{-} + 2 \,\nu^2 \,\big(-50069 + 11199 \,\delta \,\kappa_{-} - 41502 \,\kappa_{+}\big) - 10713 \,\kappa_{+}\nl
& + 39720 \,\nu^3 \,\big(2 + \,\kappa_{+}\big) + \,\nu \,\big(18530 - 32208 \,\delta \,\kappa_{-} + 53634 \,\kappa_{+}\big)\big] \,v^6\big\} \,(n\Sigma)^2\nl
& + \,\big\{-20 \,\big[18940 + 8466 \,\delta \,\kappa_{-} + 7008 \,\kappa_{+} + 10680 \,\nu^2 \,\big(2 + \,\kappa_{+}\big) - 3 \,\nu \,\big(22146 + 2281 \,\delta \,\kappa_{-} \nl
&+ 10033 \,\kappa_{+}\big)\big] \,(nv)^6 + 12 \,\big[73878 + 24257 \,\delta \,\kappa_{-} + 23513 \,\kappa_{+} + 44080 \,\nu^2 \,\big(2 + \,\kappa_{+}\big) \nl
&-  \,\nu \,\big(231290 + 20306 \,\delta \,\kappa_{-} + 101717 \,\kappa_{+}\big)\big] \,(nv)^4 \,v^2 - 12 \,\big[33624 \,\nu^2 \,\big(2 + \,\kappa_{+}\big) \nl
&+ 2 \,\big(29548 + 5583 \,\delta \,\kappa_{-} + 7800 \,\kappa_{+}\big) -  \,\nu \,\big(160250 + 9813 \,\delta \,\kappa_{-} + 63483 \,\kappa_{+}\big)\big] \,(nv)^2 \,v^4 \nl
&+ 4 \,\big[48002 + 2721 \,\delta \,\kappa_{-} + 11397 \,\kappa_{+} + 22296 \,\nu^2 \,\big(2 + \,\kappa_{+}\big) -  \,\nu \,\big(113566 + 2568 \,\delta \,\kappa_{-}\nl
& + 34857 \,\kappa_{+}\big)\big] \,v^6\big\} \,\bS^2 + \,\Big(-20 \,\big\{6 \,\big(243 - 1768 \,\nu + 2782 \,\nu^2\big) \,\kappa_{-} + \,\delta \,\big[39345 - 1458 \,\kappa_{+} \nl
&+ 10680 \,\nu^2 \,\big(2 + \,\kappa_{+}\big) - 38 \,\nu \,\big(3245 + 612 \,\kappa_{+}\big)\big]\big\} \,(nv)^6 + 12 \,\big\{\,\big(744 - 15617 \,\nu + 37144 \,\nu^2\big) \,\kappa_{-} \nl
&+ \,\delta \,\big[134379 - 744 \,\kappa_{+} + 44080 \,\nu^2 \,\big(2 + \,\kappa_{+}\big) -  \,\nu \,\big(409370 + 81411 \,\kappa_{+}\big)\big]\big\} \,(nv)^4 \,v^2\nl
& - 12 \,\big\{6 \,\big(-739 + 1501 \,\nu + 938 \,\nu^2\big) \,\kappa_{-} + \,\delta \,\big[88237 + 4434 \,\kappa_{+} + 33624 \,\nu^2 \,\big(2 + \,\kappa_{+}\big)\nl
& - 6 \,\nu \,\big(41937 + 8945 \,\kappa_{+}\big)\big]\big\} \,(nv)^2 \,v^4 + 4 \,\big\{-3 \,\big(2892 - 7135 \,\nu + 4008 \,\nu^2\big) \,\kappa_{-}\nl
& + \,\delta \,\big[55499 + 8676 \,\kappa_{+} + 22296 \,\nu^2 \,\big(2 + \,\kappa_{+}\big) -  \,\nu \,\big(137578 + 32289 \,\kappa_{+}\big)\big]\big\} \,v^6\Big) \,(S\Sigma)\nl
& + \,\big\{20 \,\big[-23735 - 729 \,\delta \,\kappa_{-} + 729 \,\kappa_{+} + 10680 \,\nu^3 \,\big(2 + \,\kappa_{+}\big) + \,\nu \,\big(85325 - 3162 \,\delta \,\kappa_{-} + 1704 \,\kappa_{+}\big) \nl
&-  \,\nu^2 \,\big(181780 + 1503 \,\delta \,\kappa_{-} + 21753 \,\kappa_{+}\big)\big] \,(nv)^6 - 6 \,\big[2 \,\nu^2 \,\big(-586820 + 1734 \,\delta \,\kappa_{-}\nl
& - 83145 \,\kappa_{+}\big) - 12 \,\big(11831 + 62 \,\delta \,\kappa_{-} - 62 \,\kappa_{+}\big) + 88160 \,\nu^3 \,\big(2 + \,\kappa_{+}\big) \nl
&+ \,\nu \,\big(563466 - 32897 \,\delta \,\kappa_{-} + 31409 \,\kappa_{+}\big)\big] \,(nv)^4 \,v^2 + 12 \,\big[-35720 + 2217 \,\delta \,\kappa_{-}\nl
& + \,\nu^2 \,\big(-340023 + 6999 \,\delta \,\kappa_{-} - 60669 \,\kappa_{+}\big) - 3 \,\nu \,\big(-56914 + 5223 \,\delta \,\kappa_{-} - 6701 \,\kappa_{+}\big)\nl
& - 2217 \,\kappa_{+} + 33624 \,\nu^3 \,\big(2 + \,\kappa_{+}\big)\big] \,(nv)^2 \,v^4 - 2 \,\big[2 \,\nu^2 \,\big(-160933 + 8580 \,\delta \,\kappa_{-} - 40869 \,\kappa_{+}\big)\nl
& + 18 \,\big(-1493 + 482 \,\delta \,\kappa_{-} - 482 \,\kappa_{+}\big) + 44592 \,\nu^3 \,\big(2 + \,\kappa_{+}\big) \nl
&+ \,\nu \,\big(176176 - 26847 \,\delta \,\kappa_{-} + 44199 \,\kappa_{+}\big)\big] \,v^6\big\} \,\bSigma^2\,,\nl
f_8^1&=\,\big\{4 \,\big[273931 - 90475 \,\delta \,\kappa_{-} + \,\nu \,\big(-101017 + 85054 \,\delta \,\kappa_{-} - 190731 \,\kappa_{+}\big) + 260591 \,\kappa_{+} \nl
&+ 14 \,\nu^2 \,\big(2 + \,\kappa_{+}\big)\big] \,(nv)^2 + 4 \,\big[-89163 + 14451 \,\delta \,\kappa_{-} - 83310 \,\kappa_{+} + 9122 \,\nu^2 \,\big(2 + \,\kappa_{+}\big) \nl
&+ \,\nu \,\big(-17229 - 13020 \,\delta \,\kappa_{-} + 14347 \,\kappa_{+}\big)\big] \,v^2\big\} \,(vS)^2\nl
& + \,\Big(4 \,\big\{\,\big(-351066 + 637685 \,\nu - 340230 \,\nu^2\big) \,\kappa_{-} \nl
&+ \,\delta \,\big[-165796 + \,\nu \,\big(159708 - 275785 \,\kappa_{+}\big) + 351066 \,\kappa_{+} + 14 \,\nu^2 \,\big(2 + \,\kappa_{+}\big)\big]\big\} \,(nv)^2 \nl
&+ 4 \,\big\{\,\big(97761 - 85171 \,\nu + 42958 \,\nu^2\big) \,\kappa_{-} + \,\delta \,\big[11030 - 97761 \,\kappa_{+} + 9122 \,\nu^2 \,\big(2 + \,\kappa_{+}\big) \nl
&+ \,\nu \,\big(-92446 + 27367 \,\kappa_{+}\big)\big]\big\} \,v^2\Big) \,(vS) \,(v\Sigma) + \,\big\{-2 \,\big[570262 + 351066 \,\delta \,\kappa_{-} \nl
&- 3 \,\nu \,\big(507492 + 152245 \,\delta \,\kappa_{-} - 386289 \,\kappa_{+}\big) + 2 \,\nu^2 \,\big(420201 + 85061 \,\delta \,\kappa_{-} - 360846 \,\kappa_{+}\big)\nl
&- 351066 \,\kappa_{+} + 28 \,\nu^3 \,\big(2 + \,\kappa_{+}\big)\big] \,(nv)^2 - 2 \,\big[-125332 - 97761 \,\delta \,\kappa_{-} \nl
&+ \,\nu \,\big(305810 + 56269 \,\delta \,\kappa_{-} - 251791 \,\kappa_{+}\big) - 2 \,\nu^2 \,\big(171541 + 8459 \,\delta \,\kappa_{-} - 35826 \,\kappa_{+}\big) \nl
&+ 97761 \,\kappa_{+} + 18244 \,\nu^3 \,\big(2 + \,\kappa_{+}\big)\big] \,v^2\big\} \,(v\Sigma)^2 + \,\big\{2 \,\big[-3056658 + 929364 \,\delta \,\kappa_{-}\nl
& - 2218433 \,\kappa_{+} + 46079 \,\nu^2 \,\big(2 + \,\kappa_{+}\big) + \,\nu \,\big(1676740 - 620106 \,\delta \,\kappa_{-} + 1491545 \,\kappa_{+}\big)\big] \,(nv)^3 \nl
&+ \,\big[3214028 - 874916 \,\delta \,\kappa_{-} + \,\nu \,\big(361732 + 302654 \,\delta \,\kappa_{-} - 80076 \,\kappa_{+}\big) + 2154322 \,\kappa_{+} \nl
&- 475778 \,\nu^2 \,\big(2 + \,\kappa_{+}\big)\big] \,(nv) \,v^2\big\} \,(vS) \,(nS) + \,\Big(\,\big\{\,\big(3147797 - 5829107 \,\nu + 2434345 \,\nu^2\big) \,\kappa_{-}\nl
& + \,\delta \,\big[-801874 - 3147797 \,\kappa_{+} + 46079 \,\nu^2 \,\big(2 + \,\kappa_{+}\big) + \,\nu \,\big(1009158 + 2111651 \,\kappa_{+}\big)\big]\big\} \,(nv)^3 \nl
&+ \,\big\{\,\big(-1514619 + 1941197 \,\nu - 367419 \,\nu^2\big) \,\kappa_{-} + \,\delta \,\big[997862 + \,\nu \,\big(114558 - 191365 \,\kappa_{+}\big)\nl
& + 1514619 \,\kappa_{+} - 237889 \,\nu^2 \,\big(2 + \,\kappa_{+}\big)\big]\big\} \,(nv) \,v^2\Big) \,(v\Sigma) \,(nS) \nl
&+ \,\big\{2 \,\big[5227548 - 1216896 \,\delta \,\kappa_{-} + \,\nu \,\big(-6893056 + 684456 \,\delta \,\kappa_{-} - 3387479 \,\kappa_{+}\big)\nl
& + 3075029 \,\kappa_{+} + 413377 \,\nu^2 \,\big(2 + \,\kappa_{+}\big)\big] \,(nv)^4 - 6 \,\big[1544838 - 328848 \,\delta \,\kappa_{-}\nl
& + \,\nu \,\big(-2011336 + 138159 \,\delta \,\kappa_{-} - 853814 \,\kappa_{+}\big) + 907817 \,\kappa_{+} + 82405 \,\nu^2 \,\big(2 + \,\kappa_{+}\big)\big] \,(nv)^2 \,v^2\nl
& + 4 \,\big[359181 - 57141 \,\delta \,\kappa_{-} + \,\nu \,\big(-616905 + 30384 \,\delta \,\kappa_{-} - 219799 \,\kappa_{+}\big) + 266628 \,\kappa_{+} \nl
&+ 30484 \,\nu^2 \,\big(2 + \,\kappa_{+}\big)\big] \,v^4\big\} \,(nS)^2 + \,\Big(\,\big\{\,\big(3147797 - 5829107 \,\nu + 2434345 \,\nu^2\big) \,\kappa_{-} \nl
&+ \,\delta \,\big[-837370 - 3147797 \,\kappa_{+} + 46079 \,\nu^2 \,\big(2 + \,\kappa_{+}\big) + \,\nu \,\big(125484 + 2111651 \,\kappa_{+}\big)\big]\big\} \,(nv)^3 \nl
&+ \,\big\{\,\big(-1514619 + 1941197 \,\nu - 367419 \,\nu^2\big) \,\kappa_{-} + \,\delta \,\big[914702 + \,\nu \,\big(700104 - 191365 \,\kappa_{+}\big) \nl
&+ 1514619 \,\kappa_{+} - 237889 \,\nu^2 \,\big(2 + \,\kappa_{+}\big)\big]\big\} \,(nv) \,v^2\Big) \,(vS) \,(n\Sigma) + \,\big\{\,\big[2673496 + 3147797 \,\delta \,\kappa_{-}\nl
& + \,\nu^2 \,\big(1386252 + 1194133 \,\delta \,\kappa_{-} - 5417435 \,\kappa_{+}\big) - 3147797 \,\kappa_{+} - 92158 \,\nu^3 \,\big(2 + \,\kappa_{+}\big)\nl
& + \,\nu \,\big(-4318238 - 3970379 \,\delta \,\kappa_{-} + 10265973 \,\kappa_{+}\big)\big] \,(nv)^3 + \,\big[-688444 - 1514619 \,\delta \,\kappa_{-} \nl
&+ 3 \,\nu \,\big(-120938 + 355427 \,\delta \,\kappa_{-} - 1365173 \,\kappa_{+}\big) + 1514619 \,\kappa_{+} + 475778 \,\nu^3 \,\big(2 + \,\kappa_{+}\big) \nl
&+ \,\nu^2 \,\big(-1396092 - 64765 \,\delta \,\kappa_{-} + 447495 \,\kappa_{+}\big)\big] \,(nv) \,v^2\big\} \,(v\Sigma) \,(n\Sigma) \nl
&+ \,\Big(2 \,\big\{\,\big(-4291925 + 8939519 \,\nu - 3151201 \,\nu^2\big) \,\kappa_{-} + \,\delta \,\big[4026508 + 4291925 \,\kappa_{+}\nl
& + 413377 \,\nu^2 \,\big(2 + \,\kappa_{+}\big) - 35 \,\nu \,\big(186627 + 116341 \,\kappa_{+}\big)\big]\big\} \,(nv)^4 \nl
&- 6 \,\big\{\,\big(-1236665 + 2307365 \,\nu - 635041 \,\nu^2\big) \,\kappa_{-} + \,\delta \,\big[1502404 + 1236665 \,\kappa_{+}\nl
& + 82405 \,\nu^2 \,\big(2 + \,\kappa_{+}\big) -  \,\nu \,\big(2117145 + 991973 \,\kappa_{+}\big)\big]\big\} \,(nv)^2 \,v^2 \nl
&+ 4 \,\big\{\,\big(-323769 + 478747 \,\nu - 152020 \,\nu^2\big) \,\kappa_{-} + \,\delta \,\big[429376 + 323769 \,\kappa_{+} + 30484 \,\nu^2 \,\big(2 + \,\kappa_{+}\big) \nl
&-  \,\nu \,\big(690749 + 250183 \,\kappa_{+}\big)\big]\big\} \,v^4\Big) \,(nS) \,(n\Sigma) + \,\big\{\,\big[\,\nu \,\big(-6095254 + 6505727 \,\delta \,\kappa_{-} - 15089577 \,\kappa_{+}\big) \nl
&- 5 \,\big(174488 + 858385 \,\delta \,\kappa_{-} - 858385 \,\kappa_{+}\big) - 826754 \,\nu^3 \,\big(2 + \,\kappa_{+}\big) \nl
&+ \,\nu^2 \,\big(12407868 - 1782289 \,\delta \,\kappa_{-} + 9926159 \,\kappa_{+}\big)\big] \,(nv)^4 + 3 \,\big[\,\nu^2 \,\big(-4525560 + 358723 \,\delta \,\kappa_{-} \nl
&- 2342669 \,\kappa_{+}\big) + 85 \,\big(-980 + 14549 \,\delta \,\kappa_{-} - 14549 \,\kappa_{+}\big) + 164810 \,\nu^3 \,\big(2 + \,\kappa_{+}\big) \nl
&+ \,\nu \,\big(3418150 - 1649669 \,\delta \,\kappa_{-} + 4122999 \,\kappa_{+}\big)\big] \,(nv)^2 \,v^2 - 2 \,\big[-26330 + 323769 \,\delta \,\kappa_{-} \nl
&+ 2 \,\nu^2 \,\big(-777095 + 45626 \,\delta \,\kappa_{-} - 295809 \,\kappa_{+}\big) - 323769 \,\kappa_{+} + 60968 \,\nu^3 \,\big(2 + \,\kappa_{+}\big)\nl
& + \,\nu \,\big(1204762 - 364465 \,\delta \,\kappa_{-} + 1012003 \,\kappa_{+}\big)\big] \,v^4\big\} \,(n\Sigma)^2 + \,\big\{-4 \,\big[433968 - 47922 \,\delta \,\kappa_{-}\nl
& + \,\nu \,\big(-715366 + 10725 \,\delta \,\kappa_{-} - 315989 \,\kappa_{+}\big) + 142766 \,\kappa_{+} + 76576 \,\nu^2 \,\big(2 + \,\kappa_{+}\big)\big] \,(nv)^4\nl
& + 4 \,\big[645648 - 61356 \,\delta \,\kappa_{-} + \,\nu \,\big(-789234 + 15507 \,\delta \,\kappa_{-} - 356657 \,\kappa_{+}\big) + 187518 \,\kappa_{+} \nl
&+ 80846 \,\nu^2 \,\big(2 + \,\kappa_{+}\big)\big] \,(nv)^2 \,v^2 - 8 \,\big[125822 - 7115 \,\delta \,\kappa_{-} + \,\nu \,\big(-74954 + 2894 \,\delta \,\kappa_{-} \nl
&- 34242 \,\kappa_{+}\big) + 30553 \,\kappa_{+} + 6601 \,\nu^2 \,\big(2 + \,\kappa_{+}\big)\big] \,v^4\big\} \,\bS^2\nl
& + \,\Big(-8 \,\big\{\,\big(-95344 + 259201 \,\nu - 59738 \,\nu^2\big) \,\kappa_{-} + \,\delta \,\big[354179 + 95344 \,\kappa_{+} + 38288 \,\nu^2 \,\big(2 + \,\kappa_{+}\big)\nl
& -  \,\nu \,\big(485847 + 163357 \,\kappa_{+}\big)\big]\big\} \,(nv)^4 + 8 \,\big\{\,\big(-124437 + 308794 \,\nu - 71437 \,\nu^2\big) \,\kappa_{-} \nl
&+ \,\delta \,\big[510814 + 124437 \,\kappa_{+} + 40423 \,\nu^2 \,\big(2 + \,\kappa_{+}\big) -  \,\nu \,\big(564209 + 186082 \,\kappa_{+}\big)\big]\big\} \,(nv)^2 \,v^2 \nl
&- 8 \,\big\{\,\big(-37668 + 65596 \,\nu - 18177 \,\nu^2\big) \,\kappa_{-} + \,\delta \,\big[170713 + 37668 \,\kappa_{+} + 6601 \,\nu^2 \,\big(2 + \,\kappa_{+}\big)\nl
& - 22 \,\nu \,\big(5271 + 1688 \,\kappa_{+}\big)\big]\big\} \,v^4\Big) \,(S\Sigma) + \,\Big(4 \,\big[\,\nu^2 \,\big(-1257006 + 49013 \,\delta \,\kappa_{-} - 375727 \,\kappa_{+}\big)\nl
& + 16 \,\big(-12070 + 5959 \,\delta \,\kappa_{-} - 5959 \,\kappa_{+}\big) + 76576 \,\nu^3 \,\big(2 + \,\kappa_{+}\big) + \,\nu \,\big(1115375 - 211279 \,\delta \,\kappa_{-} \nl
&+ 401967 \,\kappa_{+}\big)\big] \,(nv)^4 - 4 \,\big[-298355 + 124437 \,\delta \,\kappa_{-} + \,\nu^2 \,\big(-1503589 + 55930 \,\delta \,\kappa_{-} - 428094 \,\kappa_{+}\big)\nl
& - 124437 \,\kappa_{+} + 80846 \,\nu^3 \,\big(2 + \,\kappa_{+}\big) + \,\nu \,\big(1600063 - 247438 \,\delta \,\kappa_{-} + 496312 \,\kappa_{+}\big)\big] \,(nv)^2 \,v^2\nl
& + 4 \,\big\{-87541 + 37668 \,\delta \,\kappa_{-} - 6 \,\nu \,\big(-80620 + 8561 \,\delta \,\kappa_{-} - 21117 \,\kappa_{+}\big) - 37668 \,\kappa_{+}\nl
& + 13202 \,\nu^3 \,\big(2 + \,\kappa_{+}\big) + \,\nu^2 \,\big[12389 \,\delta \,\kappa_{-} - 9 \,\big(35739 + 9629 \,\kappa_{+}\big)\big]\big\} \,v^4\Big) \,\bSigma^2\,,\nl
f_8^2&=-2 \,\big[-902544 + 223199 \,\delta \,\kappa_{-} - 463094 \,\kappa_{+} + 9198 \,\nu^2 \,\big(2 + \,\kappa_{+}\big) \nl
&+ \,\nu \,\big(-396690 - 754 \,\delta \,\kappa_{-} + 5787 \,\kappa_{+}\big)\big] \,(vS)^2 - 2 \,\big\{\,\big(686293 - 899337 \,\nu - 6182 \,\nu^2\big) \,\kappa_{-} \nl
&+ \,\delta \,\big[-480890 - 686293 \,\kappa_{+} + 9198 \,\nu^2 \,\big(2 + \,\kappa_{+}\big) + \,\nu \,\big(-419408 + 6541 \,\kappa_{+}\big)\big]\big\} \,(vS) \,(v\Sigma)\nl
& + \,\big[-386486 - 686293 \,\delta \,\kappa_{-} + \,\nu \,\big(96445 + 452939 \,\delta \,\kappa_{-} - 1825525 \,\kappa_{+}\big) + 686293 \,\kappa_{+} \nl
&+ 18396 \,\nu^3 \,\big(2 + \,\kappa_{+}\big) + \,\nu^2 \,\big(-910930 + 7690 \,\delta \,\kappa_{-} + 5392 \,\kappa_{+}\big)\big] \,(v\Sigma)^2 \nl
&+ \,\big[-6363300 + 2012530 \,\delta \,\kappa_{-} - 3332386 \,\kappa_{+} + 140100 \,\nu^2 \,\big(2 + \,\kappa_{+}\big) \nl
&- 2 \,\nu \,\big(1950318 + 7900 \,\delta \,\kappa_{-} + 169755 \,\kappa_{+}\big)\big] \,(nv) \,(vS) \,(nS) \nl
&+ \,\big\{\,\big(2672458 - 3863205 \,\nu - 38450 \,\nu^2\big) \,\kappa_{-} + \,\delta \,\big[-2063276 - 2672458 \,\kappa_{+} \nl
&+ 70050 \,\nu^2 \,\big(2 + \,\kappa_{+}\big) -  \,\nu \,\big(1928546 + 161855 \,\kappa_{+}\big)\big]\big\} \,(nv) \,(v\Sigma) \,(nS)\nl
& + \,\big\{2 \,\big[7031031 - 1519937 \,\delta \,\kappa_{-} + 2903969 \,\kappa_{+} - 3306 \,\nu^2 \,\big(2 + \,\kappa_{+}\big)\nl
& + \,\nu \,\big(3359898 + 12907 \,\delta \,\kappa_{-} + 727107 \,\kappa_{+}\big)\big] \,(nv)^2 + \,\big[-9180930 + 1353406 \,\delta \,\kappa_{-}\nl
& - 3337204 \,\kappa_{+} + 40356 \,\nu^2 \,\big(2 + \,\kappa_{+}\big) - 2 \,\nu \,\big(1918032 + 4963 \,\delta \,\kappa_{-} + 612954 \,\kappa_{+}\big)\big] \,v^2\big\} \,(nS)^2\nl
& + \,\big\{\,\big(2672458 - 3863205 \,\nu - 38450 \,\nu^2\big) \,\kappa_{-} + \,\delta \,\big[70050 \,\nu^2 \,\big(2 + \,\kappa_{+}\big) \nl
&-  \,\nu \,\big(2054114 + 161855 \,\kappa_{+}\big) - 2 \,\big(908446 + 1336229 \,\kappa_{+}\big)\big]\big\} \,(nv) \,(vS) \,(n\Sigma) \nl
&+ \,\big\{817592 + 2672458 \,\delta \,\kappa_{-} - 2672458 \,\kappa_{+} - 140100 \,\nu^3 \,\big(2 + \,\kappa_{+}\big) \nl
&+ \,\nu \,\big(521708 - 1850675 \,\delta \,\kappa_{-} + 7195591 \,\kappa_{+}\big) + \,\nu^2 \,\big[-54250 \,\delta \,\kappa_{-} \nl
&+ 44 \,\big(95009 + 8590 \,\kappa_{+}\big)\big]\big\} \,(nv) \,(v\Sigma) \,(n\Sigma) + \,\Big(\,\big\{-4 \,\big(2211953 - 2682774 \,\nu + 24161 \,\nu^2\big) \,\kappa_{-} \nl
&+ 2 \,\delta \,\big[6095633 + 4423906 \,\kappa_{+} - 3306 \,\nu^2 \,\big(2 + \,\kappa_{+}\big) + \,\nu \,\big(3369929 + 714200 \,\kappa_{+}\big)\big]\big\} \,(nv)^2 \nl
&+ \,\big\{-2 \,\big(-2345305 + 2098821 \,\nu + 326 \,\nu^2\big) \,\kappa_{-} + 2 \,\delta \,\big[-4610579 - 2345305 \,\kappa_{+} \nl
&+ 20178 \,\nu^2 \,\big(2 + \,\kappa_{+}\big) -  \,\nu \,\big(1887473 + 607991 \,\kappa_{+}\big)\big]\big\} \,v^2\Big) \,(nS) \,(n\Sigma)\nl
& + \,\big\{\,\big[529810 - 4423906 \,\delta \,\kappa_{-} + \,\nu \,\big(-9756929 + 2325674 \,\delta \,\kappa_{-} - 11173486 \,\kappa_{+}\big) \nl
&+ 4423906 \,\kappa_{+} + 6612 \,\nu^3 \,\big(2 + \,\kappa_{+}\big) - 2 \,\nu^2 \,\big(3364415 + 11254 \,\delta \,\kappa_{-} + 702946 \,\kappa_{+}\big)\big] \,(nv)^2\nl
& + \,\big[5 \,\big(-185396 + 469061 \,\delta \,\kappa_{-} - 469061 \,\kappa_{+}\big) - 40356 \,\nu^3 \,\big(2 + \,\kappa_{+}\big) \nl
&+ \,\nu^2 \,\big(3647998 - 10252 \,\delta \,\kappa_{-} + 1226234 \,\kappa_{+}\big) + \,\nu \,\big(9257888 - 745415 \,\delta \,\kappa_{-}\nl
& + 5436025 \,\kappa_{+}\big)\big] \,v^2\big\} \,(n\Sigma)^2 + \,\big\{-2 \,\big[759660 - 171224 \,\delta \,\kappa_{-} + 1669 \,\nu \,\delta \,\kappa_{-} \nl
&+ 412592 \,\kappa_{+} + 22248 \,\nu^2 \,\big(2 + \,\kappa_{+}\big) + 24 \,\nu \,\big(4621 + 8133 \,\kappa_{+}\big)\big] \,(nv)^2\nl
& + \,\big[1406688 - 302336 \,\delta \,\kappa_{-} + 803672 \,\kappa_{+} - 7320 \,\nu^2 \,\big(2 + \,\kappa_{+}\big) + \,\nu \,\big(417852 + 2806 \,\delta \,\kappa_{-} \nl
&+ 444750 \,\kappa_{+}\big)\big] \,v^2\big\} \,\bS^2 + \,\Big(-2 \,\big\{\,\big(-583816 + 491373 \,\nu - 28924 \,\nu^2\big) \,\kappa_{-}\nl
& + \,\delta \,\big[976976 + 583816 \,\kappa_{+} + 22248 \,\nu^2 \,\big(2 + \,\kappa_{+}\big) + \,\nu \,\big(80996 + 193523 \,\kappa_{+}\big)\big]\big\} \,(nv)^2\nl
& - 4 \,\big\{2 \,\big(138251 - 95925 \,\nu + 488 \,\nu^2\big) \,\kappa_{-} + \,\delta \,\big[-499190 - 276502 \,\kappa_{+} + 1830 \,\nu^2 \,\big(2 + \,\kappa_{+}\big) \nl
&-  \,\nu \,\big(87269 + 110486 \,\kappa_{+}\big)\big]\big\} \,v^2\Big) \,(S\Sigma) + \,\big\{\,\big[-118138 + 583816 \,\delta \,\kappa_{-} \nl
&- 583816 \,\kappa_{+} + 44496 \,\nu^3 \,\big(2 + \,\kappa_{+}\big) + \,\nu^2 \,\big(34756 + 25586 \,\delta \,\kappa_{-} + 361460 \,\kappa_{+}\big)\nl
& + \,\nu \,\big(2720885 - 148925 \,\delta \,\kappa_{-} + 1316557 \,\kappa_{+}\big)\big] \,(nv)^2 + \,\big[357386 - 553004 \,\delta \,\kappa_{-}\nl
& + \,\nu \,\big(-2888077 + 81364 \,\delta \,\kappa_{-} - 1187372 \,\kappa_{+}\big) + \,\nu^2 \,\big(-225662 + 854 \,\delta \,\kappa_{-} - 442798 \,\kappa_{+}\big)\nl
& + 553004 \,\kappa_{+} + 7320 \,\nu^3 \,\big(2 + \,\kappa_{+}\big)\big] \,v^2\big\} \,\bSigma^2\,,\nl
f_8^3&=3 \,\big[11798 + 265 \,\delta \,\kappa_{-} + 4 \,\nu \,\big(-10242 + 463 \,\delta \,\kappa_{-} - 5019 \,\kappa_{+}\big) + 4977 \,\kappa_{+} + 672 \,\nu^2 \,\big(2 + \,\kappa_{+}\big)\big] \,(nS)^2\nl
& + 6 \,\Big\{-2 \,\big(1178 - 5217 \,\nu + 2020 \,\nu^2\big) \,\kappa_{-} + \,\delta \,\big[4643 + 2356 \,\kappa_{+} + 336 \,\nu^2 \,\big(2 + \,\kappa_{+}\big)\nl
& - 2 \,\nu \,\big(10129 + 5482 \,\kappa_{+}\big)\big]\Big\} \,(nS) \,(n\Sigma) - 3 \,\big[4 \,\nu^2 \,\big(-10066 + 547 \,\delta \,\kappa_{-} - 6029 \,\kappa_{+}\big) \nl
&+ 4 \,\big(-2 + 589 \,\delta \,\kappa_{-} - 589 \,\kappa_{+}\big) + 672 \,\nu^3 \,\big(2 + \,\kappa_{+}\big) + \,\nu \,\big(10014 - 10699 \,\delta \,\kappa_{-} + 15411 \,\kappa_{+}\big)\big] \,(n\Sigma)^2 \nl
&+ \,\big[-14234 - 265 \,\delta \,\kappa_{-} - 4 \,\nu \,\big(-9402 + 463 \,\delta \,\kappa_{-} - 5019 \,\kappa_{+}\big) - 4977 \,\kappa_{+} - 672 \,\nu^2 \,\big(2 + \,\kappa_{+}\big)\big] \,\bS^2 \nl
&+ \,\Big\{4 \,\big(1178 - 5217 \,\nu + 2020 \,\nu^2\big) \,\kappa_{-} - 2 \,\delta \,\big[3671 + 2356 \,\kappa_{+} + 336 \,\nu^2 \,\big(2 + \,\kappa_{+}\big) \nl
&- 2 \,\nu \,\big(9751 + 5482 \,\kappa_{+}\big)\big]\Big\} \,(S\Sigma) + \,\big[4 \,\nu^2 \,\big(-10234 + 547 \,\delta \,\kappa_{-} - 6029 \,\kappa_{+}\big)\nl
& + 4 \,\big(13 + 589 \,\delta \,\kappa_{-} - 589 \,\kappa_{+}\big) + 672 \,\nu^3 \,\big(2 + \,\kappa_{+}\big) + \,\nu \,\big(9942 - 10699 \,\delta \,\kappa_{-} + 15411 \,\kappa_{+}\big)\big] \,\bSigma^2\,.\nn
\end{align}
\endgroup
\end{widetext}

\bibliography{refn2ss}

\end{document}